\documentclass[twocolumn]{aastex62}
\usepackage{amsmath}

\received{xxx}
\revised{yyy}
\accepted{zzz}
\submitjournal{ApJ}

\shorttitle{Radiative transfer with {\sc POLARIS}. II.}
\shortauthors{Reissl et al.}

\begin{document}

\title{Radiative transfer with {\sc POLARIS}. II.: Modeling of synthetic Galactic synchrotron observations}

\correspondingauthor{Stefan Reissl}
\email{reissl@uni-heidelberg.de}
\affil{Zentrum f{\"u}r Astronomie der Universit{\"a}t Heidelberg, Institute of Theoretical Astrophysics,\\ Albert-Ueberle-Str. 2, 69120 Heidelberg, Germany}
\author{Stefan Reissl}
\affil{Zentrum f{\"u}r Astronomie der Universit{\"a}t Heidelberg, Institute of Theoretical Astrophysics,\\ Albert-Ueberle-Str. 2, 69120 Heidelberg, Germany}

\author{Robert Brauer}
\affil{CEA Saclay - DRF/IRFU/SAp, Orme des Merisiers, B$\hat{a}$t 709, 91191 Gif sur Yvette, France}

\author{Ralf S. Klessen}
\affiliation{Universit{\"a}t Heidelberg, Zentrum f{\"u}r Astronomie, Institute of Theoretical Astrophysics,\\ Albert-Ueberle-Str. 2, 69120 Heidelberg, Germany}
\affiliation{Universit{\"a}t Heidelberg, Interdisziplin{\"a}res Zentrum f{\"u}r Wissenschaftliches Rechnen, Im Neuenheimer Feld 205, 69120 Heidelberg, Germany}

\author{Eric W. Pellegrini}
\affiliation{Zentrum f{\"u}r Astronomie der Universit{\"a}t Heidelberg, Institute of Theoretical Astrophysics,\\ Albert-Ueberle-Str. 2, 69120 Heidelberg, Germany}

\begin{abstract}
We present an updated version of {\sc POLARIS}, a well established code designated for dust polarisation and line radiative transfer (RT) in arbitrary astrophysical environments. We extend the already available capabilities with a synchrotron feature for polarised emission. Here, we combine state-of-the-art solutions of the synchrotron RT coefficients with numerical methods for solving the complete system of  equations of the RT problem, including Faraday rotation (FR) as well as Faraday conversion (FC). We validate the code against Galactic and extragalactic observations by performing a statistical analysis of synthetic all-sky synchrotron maps for positions within the galaxy and for extragalactic observations. For these test scenarios we apply a model of the Milky Way based on  sophisticated magneto-hydrodynamic (MHD) simulations and population-synthesis post-processing techniques.We explore  different parameters for modeling the distribution of free electrons and for a turbulent magnetic field component. We find that a strongly fluctuating field is necessary for simulating synthetic synchrotron observations on small scales, we argue that Faraday rotation alone can account for the depolarisation of the synchrotron signal, and we discuss the importance of the observer position within the Milky Way. Altogether, we conclude that {\sc POLARIS} is a highly reliable tool for predicting synchrotron emission and polarisation, including Faraday rotation in a realistic galactic context. It can thus contribute to  better understand the results from current and future  observational missions. 
\end{abstract}


\section{Introduction}
Magnetic fields significantly influence the time evolution of Galactic structures and contribute to regulating the birth of new generations of stars. The exact role of the magnetic field in these processes remains a field of ongoing research. 

From the observational side a plethora of methods and physical effects can be exploited in order to estimate and measure the magnetic field strength and its direction. The Zeeman effect allows to determine the line-of-sight (LOS) field strength by observing the splitting of certain molecular lines \citep[][]{Crutcher1993,Crutcher1999}. Complementary dust polarisation measurements help us to estimate the perpendicular field component vie the Chandrasekhar-Fermi method \citep[][]{Chandrasekhar1953}. Additionally, dust polarisation can let us also infer the LOS projected field orientation. Considering the limitations of Zeeman observations \citep[e.g.][]{Brauer2017} and the uncertainties resulting from an incomplete understanding of grain alignment physics \citep[][]{Lazarian2007,Andersson2015} synchrotron polarisation and FR provides a complementary method to further constrain the field properties even more.

Galactic  radio astronomy is a mature discipline dating back to the early thirties of the last century when the first diffuse low-frequency radio emission from the Milky Way was discovered \citep[][]{Jansky1933}. Later, in the 50's this was identified as synchrotron emission from the interstellar medium (ISM) \citep[][]{Kiepenheuer1950A,Kiepenheuer1950B}. Since then numerous studies and observations applying synchrotron emission \citep[e.g.][]{Higdon1979,Haslam1981A,Haslam1982B,Beck2001,Strong2004A,Page2007,Kogut2007,Jaffe2010,Fauvet2012,Iacobelli2013,Planck2016XXXV} and Faraday rotation \citep[][]{Han2008,Wolleben2010,Jaffe2010,Oppermann2012,Beck2015} delivered important information about the distribution of magnetic fields in the ISM and in extragalactic sources. 

Consequently, this amount of observations especially the ones coming from the \cite{Haslam1981A} all-sky survey and the WMAP probe \citep[see][]{Page2007} triggered numerous distinct models concerning the large-scale structure of the Galactic ISM. These models cover many parameters such as electron distribution \citep{Drimmel2001,Page2007,Cordes2002}, dust and synchrotron emission \citep{Sun2008,Beck2016,Vaisala2018}, FR \citep[][]{Beck2016,Pakmor2018}, and recently predictions for the synchrotron circular polarisation \citep{Ensslin2017}.

Further input to models of the Milky Way come from numerical simulations. Thanks to the development of new algorithms and the ever increasing capabilities of modern supercomputing facilities, state-of-the-art MHD simulations provide an unprecedented level of complexity and physical fidelity. For example, the calculations of the SILCC project \citep[][]{Walch2015,Girichidis2016} describe the dynamical evolution of the magnetised multi-phase ISM in a representative region of the Galactic disc including time-dependent chemistry \citep[][]{Glover2010,Glover2012}, a prescription of star formation \citep[using sink particles, see][]{Federrath2010} and stellar feedback \citep[such as supernovae,][]{Gatto2015,Gatto2017}, or ionizing radiation \citep[][]{Baczynski2015,Peters2017,Haid2018}, as well as cosmic rays 
\citep[][]{Girichidis2016,Girichidis2018}. Similar approaches are followed by \cite{deAvillez2005}, \cite{Joung2006}, \cite{Hill2012}, \cite{Gent2013}, \cite{Gressel2013}, \cite{Hennebelle2014}, \cite{Simpson2016}, \cite{Kim2017}, or  \cite{Hennebelle2018} with different codes and various physical processes taken into account. 

Full disc simulations with realistic ISM and magnetic field parameters are less frequently discussed in the literature. They are performed either for isolated galaxies \citep[][]{Pakmor2013,Pakmor2016,Rieder2016,Koertgen2018} or for galaxies in a full cosmological context \citep[][]{Pakmor2014,Pakmor2017,Rieder2017,MartinAlvarez2018}. Furthermore, simulations of the Milky Way with a realistic multi-phase ISM including bar, bulge, disc, and halo component are presented in \cite{Sormani2018} and the influence of Parker instabilities to star-formation in strongly magnetised and self-gravitating Galactic discs are studied in \cite{Koertgen2018}. In the current study, we specifically use data from the Auriga project \citep[][]{Grand2017,Pakmor2018} extended with an high resolution electron distribution \citep[][]{ReisslA} because it provides the best combination of high numerical resolution, number of physical processes included, and realistic treatment of magnetic field evolution. 

Connecting Galactic observations with analytical models and numerical simulations requires post-processing with a proper RT scheme. This is not a trivial task. Indeed, dust emission and polarisation by scattering, photoionization, and molecular line RT is a common feature in many codes {\citep[][]{Juvela1999,Wolf1999,Whitney2002,Niccolini2001,Gordon2001,Misselt2001,Ercolano2003,Wolf2003,Steinacker2003,Juvela2003,Min2009,Whitney2011,Baes2011,Dullemond2012,Robitaille2013,Harries2014,Reissl2016}}. However, such codes often lack a proper treatment of aligned dust grains \citep{Pelkonen2009, Reissl2016, Pelkonen2017, Juvela2018} or line RT including Zeeman splitting \citep{Larsson2014,Reissl2016,Brauer2017}.

For a for publicly available RT code with synchrotron capabilities we refer to {\sc GTRANS}\footnote{https://github.com/jadexter/grtrans} \citep[][]{Dexter2016}, which solves the RT problem on a highly relativistic environments on a Kerr metric, and to the {\sc HAMMURABI}\footnote{https://sourceforge.net/p/hammurabicode/wiki/Home/} code \citep[][]{Waelkens2009}, which has been used to produce mock Galactic all-sky maps including free- free emission and ultra-high energy cosmic ray at all frequencies
in a 3D magnetic field model and electron distribution. Both codes lack the variability concerning detectors, grid geometries, and an easy handling of external MHD data. Furthermore, {\sc HAMMURABI} is highly specialised to model the Milky Way alone for an observer placed within the model. 

In turn {\sc POLARIS}\footnote{http://www1.astrophysik.uni-kiel.de/${\sim}$polaris/} \citep[][]{Reissl2016} is a well-tested RT OpenMP parallelised code working on numerous grids (adaptive octree, spherical, cylindrical, and native Voronoi). The code is completely written in C++ and provides the standard features of dust heating and polarisation by dust scattering. Beyond that, {\sc POLARIS} comes with a state-of-the art treatment of dust grain alignment physics \citep[][]{Reissl2016,Reissl2017,Seifried2018,Reissl2018} as well as line RT including the Zeeman effect \citep[][]{Brauer2016,Brauer2017,Brauer2017B}, all wrapped into a collection of supplementing python scripts for plotting, statistical analysis, and MHD data conversion. 

Driven by the observational capabilities of new telescopes, such as WMAP, Planck, VLT, ALMA, or SKA,  as well as the vastly increasing complexity of MHD simulations, there is a need for a new and versatile RT tool that is able to combine all aspects of the physics of electromagnetic waves traveling to complex media. To achieve this we add a new C++ class to {\sc POLARIS}  and connected the code to the broader framework of Galactic disc modeling.

The paper is structured as follows. In Section \ref{sect:RTProblem} we describe basic quantities of the RT problem and discuss the RT with thermal electron and cosmic ray (CR) electrons in Section \ref{sect:RTThermal} and Section \ref{sect:RTCR}, respectively. We introduce the applied Milky Way model in Section \ref{sect:TestSetup} and discuss the ways to modify this model by an additional turbulent magnetic field component in Section\ref{sect:TurbulentB} and a CR electron distribution in Section \ref{sect:CRDistribution}. In Section \ref{sect:ResultsDiscussion} we present the comparison of synthetic maps and actual observations. This includes the similarities of different profiles of the turbulent magnetic field and electron distributions of the Galactic model and the Milky Way to quantify the predictive capability of the POLARIS code. This is followed by the evaluation of Galactic all-sky maps and extragalactic observations in Section \ref{sect:ObsSynchrotron} and Section \ref{sect:ObsSynchrotron}, respectively. Finally, we summarise our results in Section \ref{sect:Summary}.

\section{The radiative transfer (RT) problem}
\label{sect:RTProblem}
The polarisation state of radiation along its path can be conveniently quantified by the four-parameter Stokes vector 
\begin{equation}
\vec{S}=\left(I,Q,U,V\right)^T\, ,
\end{equation}
where the parameter $I$ represents the total intensity, $Q$ and $U$ describe the state of linear polarisation, and $V$ is for circular polarisation. It follows from the Stokes formalism that the linearly polarised fraction of the intensity is determined by
\begin{equation}
P_{\mathrm{l}}=\sqrt{\frac{U^2+Q^2}{I^2}}\;.
\label{eq:PlDefinition}
\end{equation}
The total polarisation is defined as 
\begin{equation}
p_{\mathrm{t}}=\sqrt{U^2+Q^2+V^2}\;.
\label{eq:PtDefinition}
\end{equation}
Typically $p_{\mathrm{t}} \ll I$, while  $p_{\mathrm{t}} = I$ means totally polarised radiation. The position angle of linear polarisation $\chi$ as observed on the plane of the sky is
\begin{equation}
\chi=\frac{1}{2} \tan^{-1}\left(\frac{U}{Q} \right).
\label{eq:LinearOrientation}
\end{equation}
{\sc POLARIS} solves the RT equation in all four Stokes parameters simultaneously \citep[][]{Reissl2016}. In the most general form this problem can be expressed as \citep[e.g.][]{Martin1971, Jones1979}:
\begin{equation}
\frac{d}{d\ell}\vec{S}=-\hat{K}\vec{S}+\vec{J}\, .
\label{eq:RTProblem1}
\end{equation}
Here, $\vec{J}$ is the emissivity and the quantity $\hat{K}$ is the $4\times4$ M{\"u}ller matrix  describing the extinction and absorption, respectively. Both $\hat{K}$ as well as $\vec{J}$ are defined by the characteristic physics of radiation passing through a medium.

Dependent on the physical problem some of the coefficients can be eliminated by rotating the polarised radiation from the lab reference frame into the target frame meaning the frame of the propagation direction (see Figure \ref{fig:Sketch}). From the definition of the Stokes vector follows for the rotation matrix
\begin{equation}
\hat{R}(\varphi) =\begin{pmatrix} 1 &  0 & 0 & 0 \\  0 & \cos(2\varphi) & -\sin(2\varphi) & 0 \\ 0 & \sin(2\varphi) & \cos(2\varphi) & 0 \\ 0 & 0 & 0 & 1 \end{pmatrix}\, ,
\end{equation}
where $\varphi$ is the angle between the x-axis of the target frame and the magnetic field direction projected into the plane perpendicular to the propagation direction of the radiation (see Figure \ref{fig:Sketch}). 
Note that ${\hat{R}^{-1}(\varphi)=\hat{R}(-\varphi)}$. Consequently, {\sc POLARIS} rotates the Stokes vector into the target frame when entering each individual grid cell and back when escaping it. 

Finally, the set of Stokes RT equations reads:
\begin{equation}
\frac{\rm{d}}{d\ell}\begin{pmatrix} I\\ Q \\ U \\ V \end{pmatrix}=\begin{pmatrix} j_{\rm{I}}\\ j_{\rm{Q}} \\ 0 \\ j_{\rm{V}} \end{pmatrix}-\begin{pmatrix} 
\alpha_{\rm I} & \alpha_{\rm Q} & 0 & \alpha_{\rm V} \\  
\alpha_{\rm Q} & \alpha_{\rm I} & \kappa_{\rm V} & 0 \\ 
0 & -\kappa_{\rm V} & \alpha_{\rm I} & \kappa_{\rm Q} \\ 
\alpha_{\rm V} & 0 & -\kappa_{\rm Q} & \alpha_{\rm I} \end{pmatrix}\begin{pmatrix} I\\ Q \\ U \\ V \end{pmatrix}\, .
\label{eq:RTProblem2}
\end{equation}
Reliable computation of synchrotron emission and polarisation rests on the availability of accurate RT coefficients of absorption and emission in an ionised plasma \cite[see][for review]{Heyvaerts2013}. An exact solution of the synchrotron RT problem requires to solve integrals over modified Bessel functions \citep[see e.g.][for details]{Rybicki1979,Huang2011,Heyvaerts2013}. Because of the high computational cost of RT simulations in media with complex density and magnetic field structure, the implementation in {\sc POLARIS} follows the approach of applying fit functions approximating the integral solutions in order to increase the performance. These are highly accurate for typical ISM-like conditions and can efficiently be evaluated during the RT simulation (for the exact errors and limitations we refer to Appendices \ref{app:Error} and \ref{app:Solver}). Finally, {\sc POLARIS} solves Equation \ref{eq:RTProblem2} along a particular line of sight (LOS) by means of ray-tracing using a Runge-Kutta solver (see e.g \cite{Ober2015} and  Appendix \ref{app:Solver}). In the ray-tracing mode of {\sc POLARIS} the rays can be either parallel for an observer placed outside the grid or they start at a HEALPIX\footnote{https://healpix.jpl.nasa.gov/} sphere converging at the observer position. The later case is for simulating all-sky maps. {\sc POLARIS} uses a sub-pixeling scheme where rays are split into sub-rays as long as neighbouring rays do not pass the same cells along their individual LOS. This ensures a accurate covering of gird structures smaller than the defined detector resolution.

The individual coefficients of the emissivity vector $\vec{J}$ and the M{\"u}ller matrix $\hat{K}$ follow from the physics of radiation-electron interaction in an ionised plasma. For a comprehensive approach for accurate synchrotron RT in complex astrophysical environments, one needs to consider two different species of electrons: CR electrons and thermalised relativistic electrons \citep[][]{Jones1977,Jones1979,Heyvaerts2013,Beck2015,Pandya2016,Dexter2016}. Synchrotron intensity as well as linear and circular polarisation emerges mostly from CR electrons whereas thermal electrons dominate Faraday rotation (FR) and Faraday conversion (FC) \citep[e.g.][]{Beck2015,Ensslin2017}.

\subsection{RT with thermal electrons}
\label{sect:RTThermal}
\begin{figure}
\begin{center}
        \begin{minipage}[c]{1.0\linewidth}
                        \begin{center}
                                \includegraphics[width=0.75\textwidth]{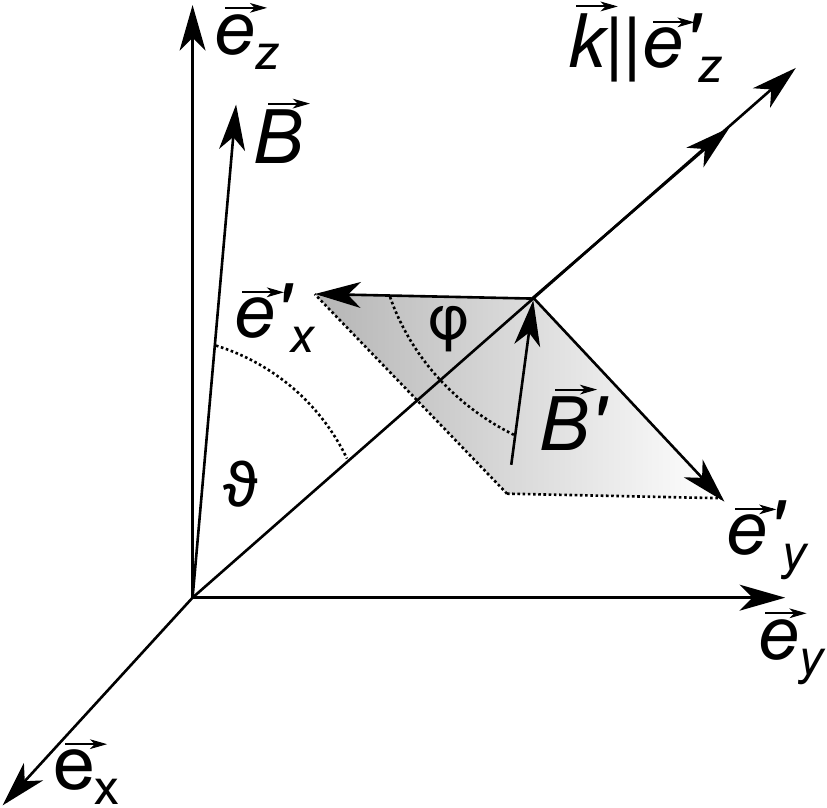}
                        \end{center}
       \end{minipage} \end{center} 
\caption{Sketch of the lab frame of reference ${(\vec{e}_{\rm x},\vec{e}_{\rm y},\vec{e}_{\rm z})}$ and the target target frame ${(\vec{e}'_{\rm x},\vec{e}'_{\rm y},\vec{e}'_{\rm z})}$ where the $\vec{e}'_{\rm x}$ corresponds to the $+Q$ Stokes parameter and $\vec{e}'_{\rm y}$ corresponds to $-Q$. The direction of light propagation $\vec{k}$ is parallel to $\vec{e}'_{\rm z}$.  The angle $\vartheta$ is defined to be between the magnetic field direction $\vec{B}$ and $\vec{k}$ whereas $\varphi$ is between $\vec{e}'_{\rm x}$ and the magnetic field vector $\vec{B}'$  projected on the $\vec{e}'_{\rm x}\vec{e}'_{\rm y}$ plane.}
\label{fig:Sketch}
\end{figure}
Thermal electrons follow a Maxwell J{\"u}ttner distribution (a relativistic Maxwellian energy distribution). In the notation of \cite{Shcherbakov2008} this distribution can be expressed with the dimensionless electron temperature
\begin{equation}
\Theta=\frac{k_{\rm B} T_{\rm e}}{m_{\rm e}c^2}
\label{eq:Theta}
\end{equation}
as parameter, where $k_{\mathrm B}$ is the Boltzmann constant,  $T_{\mathrm e}$ is the electron temperature, $c$ is the speed of light, and $m_{\mathrm e}$ is the electron mass. The Maxwell J{\"u}ttner distribution can then be written as a function of the Lorentz factor ${\gamma=(1-\beta^2)^{1/2}}$  and ${\beta=v/c}$ as
\begin{equation}
N_{\rm th}(\gamma)=\frac{n_{\rm th}\gamma^2\beta\exp\left(-\gamma/\Theta\right)}{\Theta K_{\rm 2}\left(\Theta^{-1}\right)}
\end{equation}
normalised such that ${\int N_{\rm th}(\gamma) d\gamma=n_{\rm th}}$. The quantities $n_{\mathrm th}$ and  $K_{\rm 2}(\Theta^{-1})$ are local thermal electron number density and second-order modified Bessel function, respectively. The electrons emit at a characteristic wavelength corresponding to the radius of their cyclotron orbit,
\begin{equation}
\lambda_{\rm c}=\frac{2\pi m_{\rm e}c^2}{eB}\, .
\label{eq:LambdaC}
\end{equation}
Here, $e$ is the electron charge and $B$ is the magnetic field strength. 
Examining the exact solution to this problem \citep[see e.g.][for further details]{Heyvaerts2013,Pandya2016,Dexter2016} it follows that the contribution of thermal electrons to emission and absorption is minuscule for $\Theta \ll 1$. Considering the typical ISM temperatures we assume  $j_{\mathrm I,Q,U}=0$ and $\alpha_{\mathrm I,Q,U}=0$ for our Galactic modeling.

In contrast to polarized RT with non-spherical dust grains \citep[see e.g.][]{Reissl2016} with its transfer between $Q$ and $U$ parameters, the synchrotron RT matrix $\hat{K}$ has additional coefficients that link the Stokes components $I$ and $V$. Here, we make only use of the low temperature regime with $\Theta \ll 1$, meaning $T_{\mathrm e}\ll 10^{10}\ \mathrm{K}$, which is reasonable for the ISM. Hence, the Faraday coefficients given in \cite{Huang2011} and \cite{Dexter2016} converge to
\begin{equation}
\kappa_{\rm Q}\left(\lambda,\vartheta\right) = \frac{1}{4\pi^2}\frac{n_{\rm th}e^4B^2}{m_{\rm e}^3c^6}\lambda^3 \sin^2(\vartheta)\, ,
\label{eq:kappaQ}
\end{equation}
where $\kappa_{\mathrm Q}$ is referred to as the FC coefficient and the corresponding FR coefficient is defined as
\begin{equation}
\kappa_{\rm V}\left(\lambda,\vartheta\right) =  \frac{1}{2\pi}\frac{n_{\rm th}e^2B}{m_{\rm e}^2c^4}\lambda^2 \cos(\vartheta)\, .
\label{eq:kappaV}
\end{equation}
These equations also coincide with the coefficients given in \cite{Ensslin2003}. Here, the angle $\vartheta$ is between the direction of light propagation and the magnetic field (see Fig
.\ref{fig:Sketch}).

Polarised radiation passing ionised and magnetised regions change their position angle $\chi$ (see Equation \ref{eq:LinearOrientation}) and the actually observed orientation becomes
\begin{equation}
\chi_{\rm obs}=\chi+\lambda^2 \times RM\, .
\label{eq:ChiRot}
\end{equation}
The quantity
\begin{equation}
RM = \frac{1}{2\pi}\frac{e^2}{m_{\rm e}^2c^4} \int n_{\rm th} B_{||} d\ell
\end{equation}
is the rotation measure, closely connected to the FR coefficient via $dRM = \lambda^{-2}\kappa_{\mathrm V}d\ell$ where $B_{||} = B \cos(\vartheta)$ is the LOS magnetic field component (see also Figure \ref{fig:Sketch}). 

The FR of the polarisation angle $\chi$ may have a severe impact on the observed polarisation of a synchrotron source. The Stokes Q and U components can change sign or even completely depolarise. The Faraday depolarisation DP can be quantified by 
\begin{equation}
\label{eq:depol}
DP=\frac{I_{\lambda_{1}}\times P_{\mathrm{l},\lambda_{1}}}{I_{\lambda_{2}}\times P_{\mathrm{l},\lambda_{2}}} \left(\frac{\lambda_{1}}{\lambda_{2}}\right)^{\alpha}\, ,
\end{equation}
where $P_{\mathrm{l},\lambda_{1}}$ and $P_{\mathrm{l},\lambda_{2}}$ are the polarisation fractions at any two different wavelengths $\lambda_{1}$ and $\lambda_{2}$ and where $\alpha$ is the spectral index. More precisely, $DP=1$ means no depolarisation, whereas $DP=0$ corresponds to total depolarisation. For synchrotron radiation the spectral index is directly connected to the power-law exponent $p$ (see Eq. \ref{eq:CRDistribution}) via $\alpha=(p-1)/2$. The advantage of the quantity DP is that it removes all  depolarisation effects other than FR depolarisation.

\subsection{RT with CR electrons}
\label{sect:RTCR}
Polarised synchrotron emission results from accelerated CR electrons in the presence of a magnetic field. The distribution of CR electrons is usually modeled as a power-law 
\begin{equation}
N_{\rm CR}(\gamma) \hspace{-0.5mm}=\hspace{-0.5mm} \begin{cases} n_{\rm CR}\gamma^p(p-1)\left( \gamma_{\rm min}^{p-1}-\gamma_{\rm max}^{p-1} \right) & \hspace{-3mm} \mbox{if } \gamma_{\rm min}\hspace{-0.5mm} < \gamma \hspace{-0.5mm} <  \gamma_{\rm max} \\0 & \hspace{-3mm} \mbox{otherwise}   \end{cases}\, ,
\label{eq:CRDistribution}
\end{equation}
with ${\int N_{\rm CR}(\gamma) d\gamma=n_{\rm CR}}$ and sharp cut-offs at $\gamma_{\rm min}$ and $\gamma_{\rm max}$, respectively. Here, $n_{\mathrm CR}$ is the CR electron density and $p$ is the power-law index. 
For the coefficients of emissivity and absorption we implemented approximate solutions as presented in \cite{Pandya2016} (their equations in our notation). Polarised synchrotron emission is defined by the coefficients of total emission
\begin{equation}
\begin{split}
j_{\rm I}\left(\lambda\right) = \gamma_{\rm min}^{1-p} \frac{1}{\lambda_{\rm c}}  \frac{n_{\rm CR} e^2  3^{\frac{p}{2}} (p-1)\sin(\vartheta)}{2(p+1)\left(\gamma_{\rm min}^{1-p}-\gamma_{\rm max}^{1-p}  \right)} \times \qquad\qquad \qquad \\ \qquad\Gamma\left(\frac{3p-1}{12}\right)\Gamma\left(\frac{3p+19}{12}\right)\left( \frac{\lambda_{\rm c}}{\lambda \sin(\vartheta)}  \right)^{-\frac{p-1}{2}}\, ,
\end{split}
\label{eq:CRJI}
\end{equation}
linearly polarised emission
\begin{equation}
j_{\rm Q}\left(\lambda\right) = j_{\rm I}\left(\lambda\right)\left( -\frac{p+1}{p+7/3} \right)\, ,
\end{equation}
and circularly polarised emission
\begin{equation}
j_{\rm V}\left(\lambda\right) = j_{\rm I}\left(\lambda\right) \left( -\frac{171}{250}\frac{\lambda_{\rm c}p^{\frac{1}{2}}}{3\lambda\tan(\vartheta)} \right)\, .
\end{equation}
Here $\Gamma$ is the gamma function. Tests of this approach against the exact integral solutions implemented in the {\sc SYMPHONY}\footnote{https://github.com/AFD-Illinois/{\sc SYMPHONY}} code can be found in Appendix \ref{app:Error}. Note that $j_{\rm U}$ is not required because of the rotation introduced in Section \ref{sect:RTProblem}. It follow that the maximal possible degree of linear polarisation is directly connected to the power-law index $p$ \citep[see also][]{Rybicki1979}, since
\begin{equation}
 \max\left(P_{\mathrm l}\right)= \frac{ \left| j_{\mathrm Q} \right| }{j_{\mathrm I}}=\frac{p+1}{p+7/3}\, .
\label{eq:MaxPl} 
\end{equation}
In contrast to thermal electrons, the CR electron cannot be considered in thermal equilibrium with their environment. Hence, Kirchhoff's law does not apply here. Solutions of absorption by CR electrons are derived in \cite{Pandya2016} where the coefficients for total synchrotron absorption $\alpha_{\rm I}\left(\lambda\right)$, linear polarisation $\alpha_{\rm Q}\left(\lambda\right)$, as well as circular polarisation $\alpha_{\rm V}\left(\lambda\right)$ are written as
\begin{equation}
\begin{split}
\alpha_{\rm I}\left(\lambda\right) = \gamma_{\rm min}^{1-p} \frac{\lambda n_{\rm CR} e^2}{ m_{\rm e}c^2}  \frac{  3^{\frac{p+1}{2}} (p-1)}{4\left(\gamma_{\rm min}^{1-p}-\gamma_{\rm max}^{1-p}  \right)} \times   \qquad\qquad\qquad   \\ \qquad \qquad \Gamma\left(\frac{3p+12}{12}\right)\Gamma\left(\frac{3p+22}{12}\right)\left( \frac{\lambda_{\rm c}}{\lambda \sin(\vartheta)}  \right)^{-\frac{p+2}{2}}\, ,
\end{split}
\label{eq:CRAlphaI}
\end{equation}
\begin{equation}
\alpha_{\rm Q}\left(\lambda\right) = \frac{996\ \alpha_{\rm I}\left(\lambda\right) }{1000}\left( -\frac{3(p-1)^{\frac{43}{500}}}{4}  \right)\, ,
\label{eq:CRAlphaQ}
\end{equation}
and
\begin{equation}
\begin{split}
\alpha_{\rm V}\left(\lambda\right) = \alpha_{\rm I}\left(\lambda\right)k_{\mathrm{V}}\left(\vartheta\right) \left[ -\frac{7}{4}\left(\frac{71p}{100}+\frac{22}{625}\right)^{\frac{197}{500}}  \right] \times \qquad\quad \\  \left[  \left(  \sin^{-\frac{48}{25}}(\vartheta)-1 \right)^{\frac{64}{125}} \left(\frac{\lambda_{\rm c}}{\lambda}\right)^{-\frac{1}{2}}     \right]\, ,
\end{split}
\label{eq:CRAlphaV}
\end{equation}
respectively. Here, the function $k_{\mathrm{V}}\left(\vartheta\right)$ is an additional correction to minimise the error of $\alpha_{\rm V}\left(\lambda\right)$ (see Appendix \ref{app:Error} for details). 

The Faraday mixing coefficients for CR electrons are usually considered to be irrelevant in their contributions to the RT in comparison to the coefficients of thermal electrons. Although, mathematical expressions for the FR as well as the FC coefficients exist \citep[see][e.g.]{Jones1977,Huang2011,Dexter2016} and they might contribute to the synchrotron RT (given certain conditions) we apply $\kappa_{\rm Q}\left(\lambda\right) = 0$ and $\kappa_{\rm V}\left(\lambda\right) = 0$ for the RT in this paper.

\section{Modeling of Milky Way-like galaxies}
\label{sect:TestSetup}
In order to quantify the reliability of the {\sc POLARIS} RT simulations we follow previous publications  \cite[e.g][]{Waelkens2009,Dexter2016,King2016,Ensslin2017} to compare the {\sc POLARIS} results to actual observations. As input for {\sc POLARIS} we use data from the Auriga project \citep[][]{Grand2017,ReisslA} which provides a  high-resolution cosmological MHD zoom simulations of Milky Way-like galaxies.

In particular, the Auriga galaxy Au-6 at high resolution with a halo mass of $10^{12}\,\mathrm{M_\odot}$ and a stellar mass of $6 \times 10^{10}\,\mathrm{M_\odot}$ resembles the Milky Way in many ways. It reproduces key properties of the stellar disc \citep[][]{Grand2017, Grand2018}, the gas disc \citep[][]{Marinacci2017}, the stellar halo \citep[][]{Monachesi2016}, the magnetic field structure \citep[][]{Pakmor2017}, and its satellite population \citep[][]{Simpson2018}. For synthetic observations, see also \citep[see][for details]{Grand2018A,Pakmor2018}. Recently, the Au-6 data has formed the basis for a new population synthesis model of star formation \citep[][]{ReisslA} resulting in additional physical parameters such as thermal electron density distributions and electron temperatures which allows for the calculation of synthetic emission lines, such as $H_\alpha$, $H_\beta$, $O_{\mathrm{II}}$ , $O_{\mathrm{III}}$, $[S_{\mathrm{III}}]$, etc.

We use the Auriga galaxy Au-6 a as a framework to simulate synchrotron emission and FR effects. Our approach is based on the original Au-6 density and magnetic field distribution \citep[][]{Grand2017} and the thermal electron distribution provided by \cite{ReisslA}. We also include a realistic model for the small scale structure of the magnetic field and add an additional CR electron component in a separate post-processing step as described below in order to provide the means of evaluating the new synchrotron feature of {\sc POLARIS}.
\begin{figure*}
 \vspace{-3mm}
 \begin{center}
         \begin{minipage}[c]{1.0\linewidth}
                         \begin{center}
                                 \includegraphics[width=0.49\textwidth]{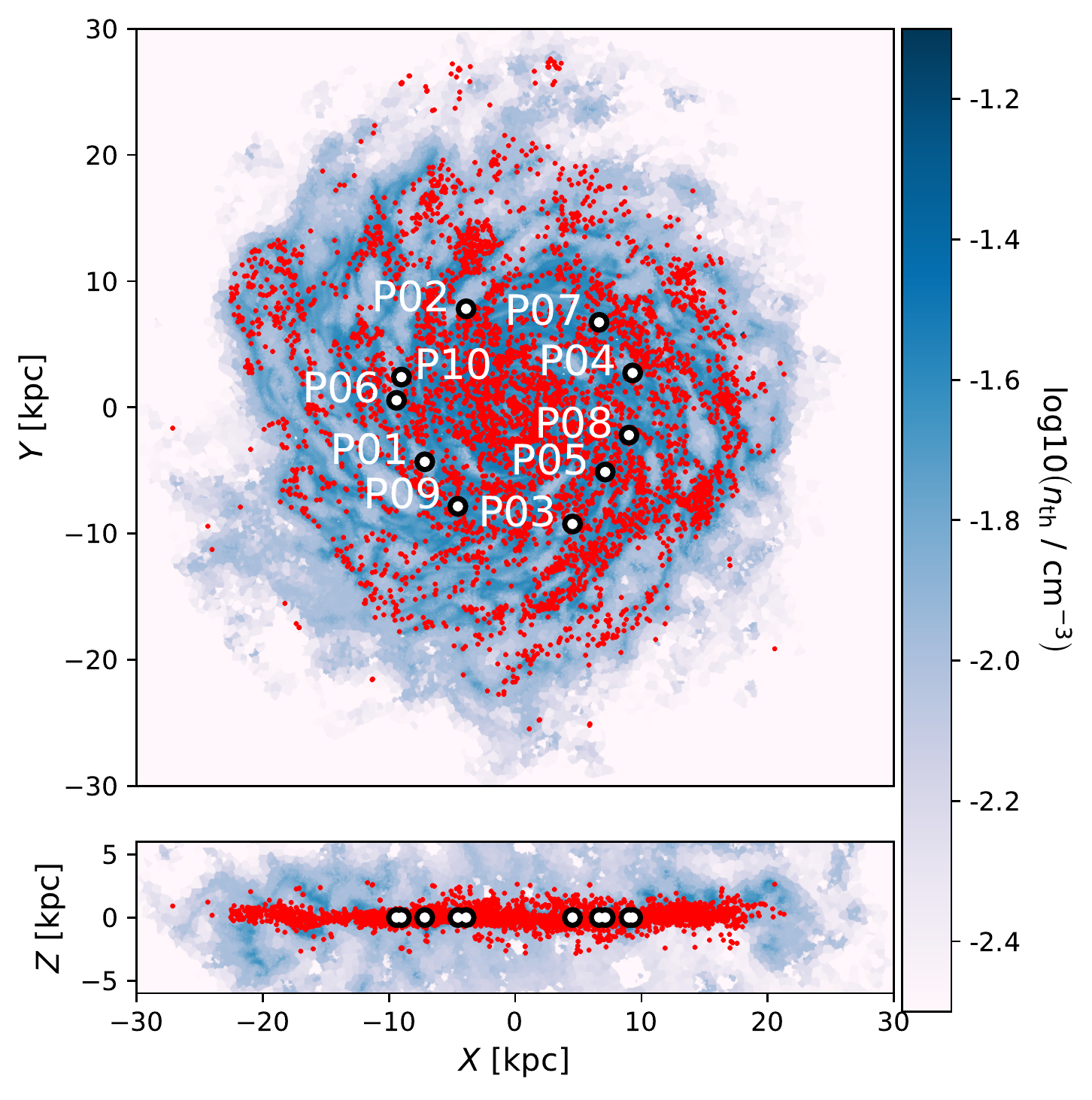}
                                 \includegraphics[width=0.49\textwidth]{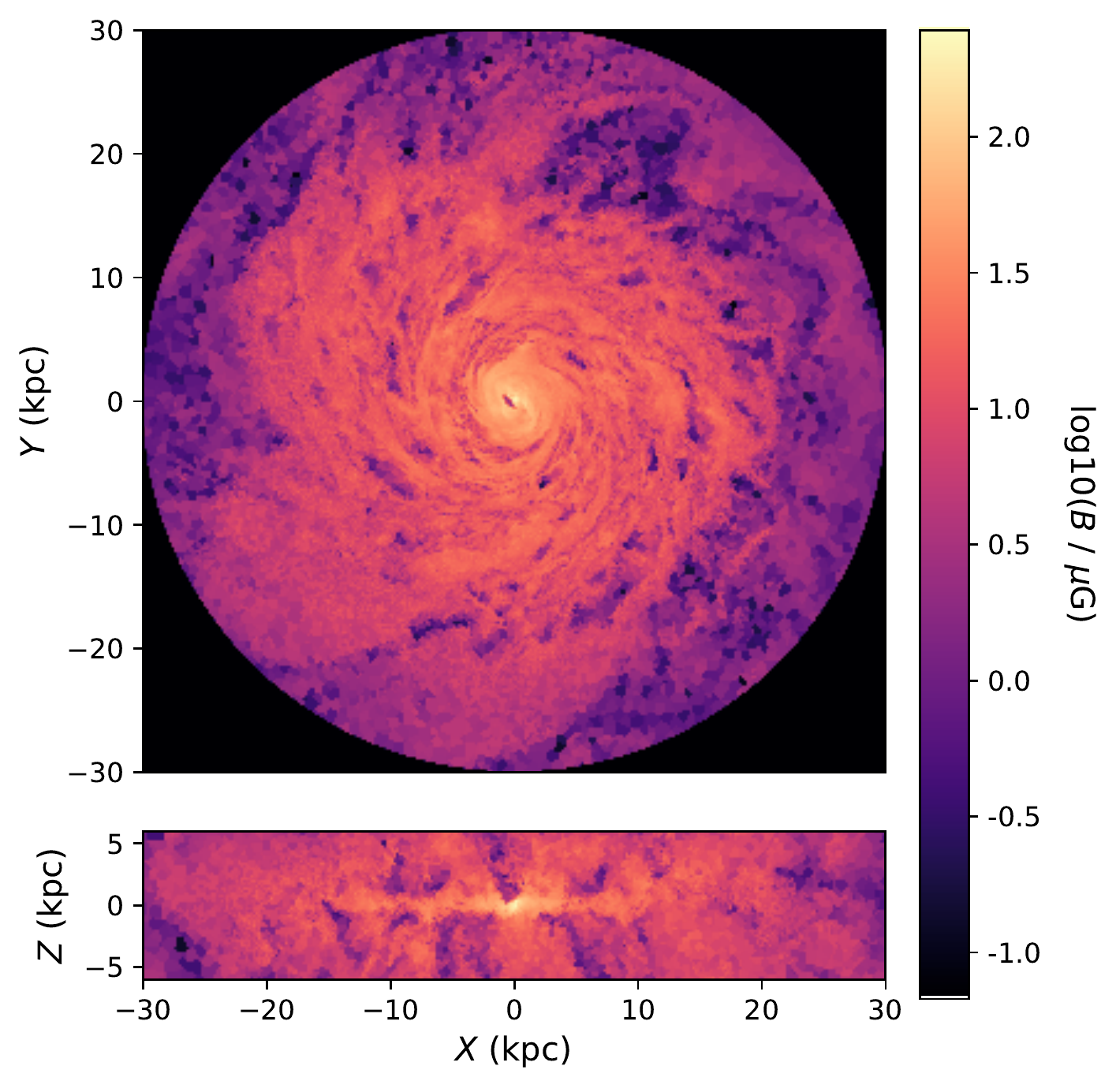}
                                 \vspace{-2mm}
                                  \caption{Left panel: Distribution of thermal electrons $n_{\mathrm{th}}$ in the disc midplane (top) and perpendicular to it (bottom) provided by the population synthesis model of \protect\cite{ReisslA}. Red dots represent the cluster population and black circles and white labels indicate the selected observer positions. Right panel: Corresponding magnetic field strength of the Au-6 galaxy \citep[][]{Grand2017} modified as outlined in Section \ref{sect:TurbulentB}.}
                                  \label{fig:MidThB}
                         \end{center}
                       \end{minipage} 
         \begin{minipage}[c]{1.0\linewidth}
                         \begin{center}
                                 \includegraphics[width=0.49\textwidth]{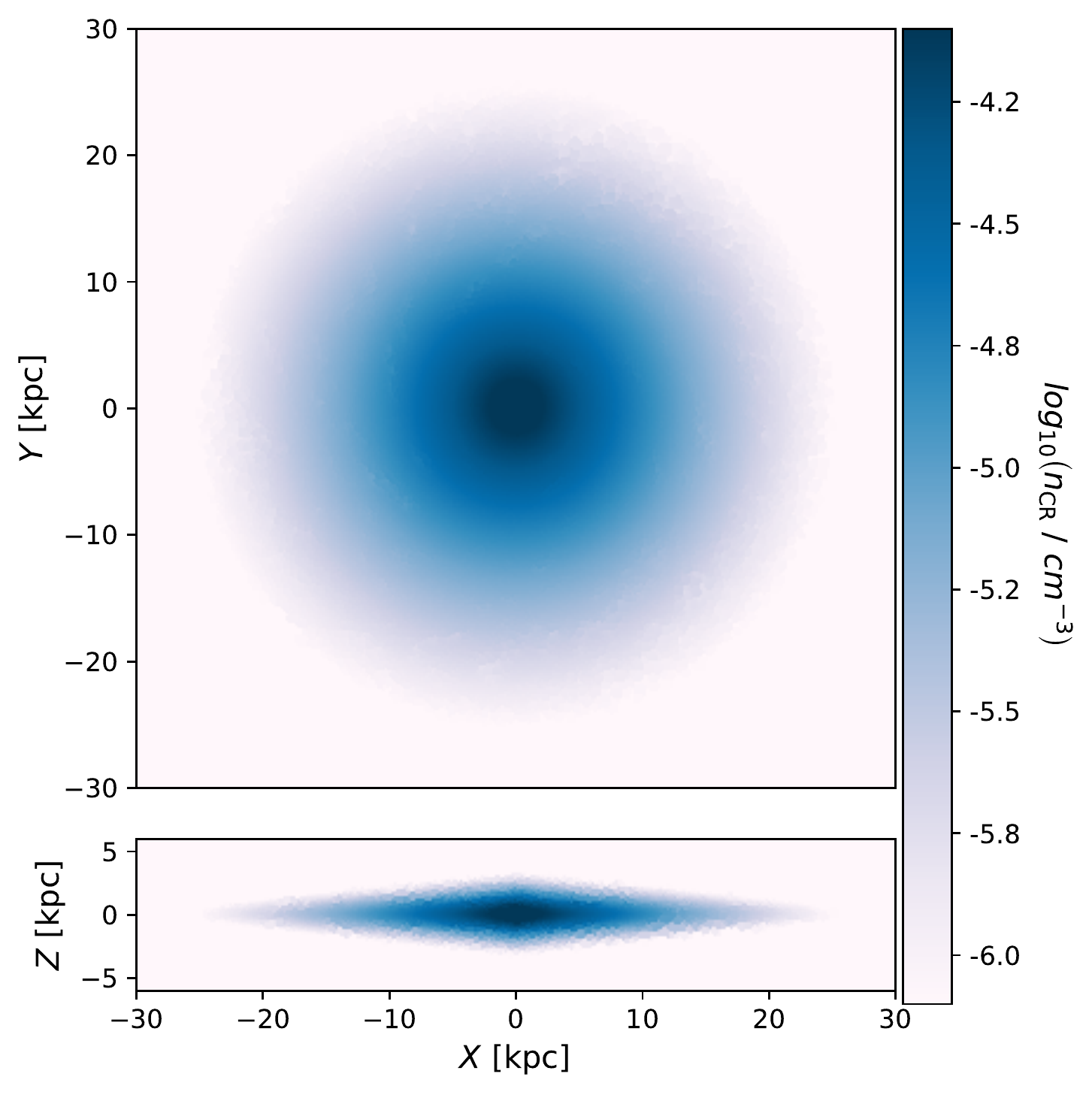}
                                 \includegraphics[width=0.49\textwidth]{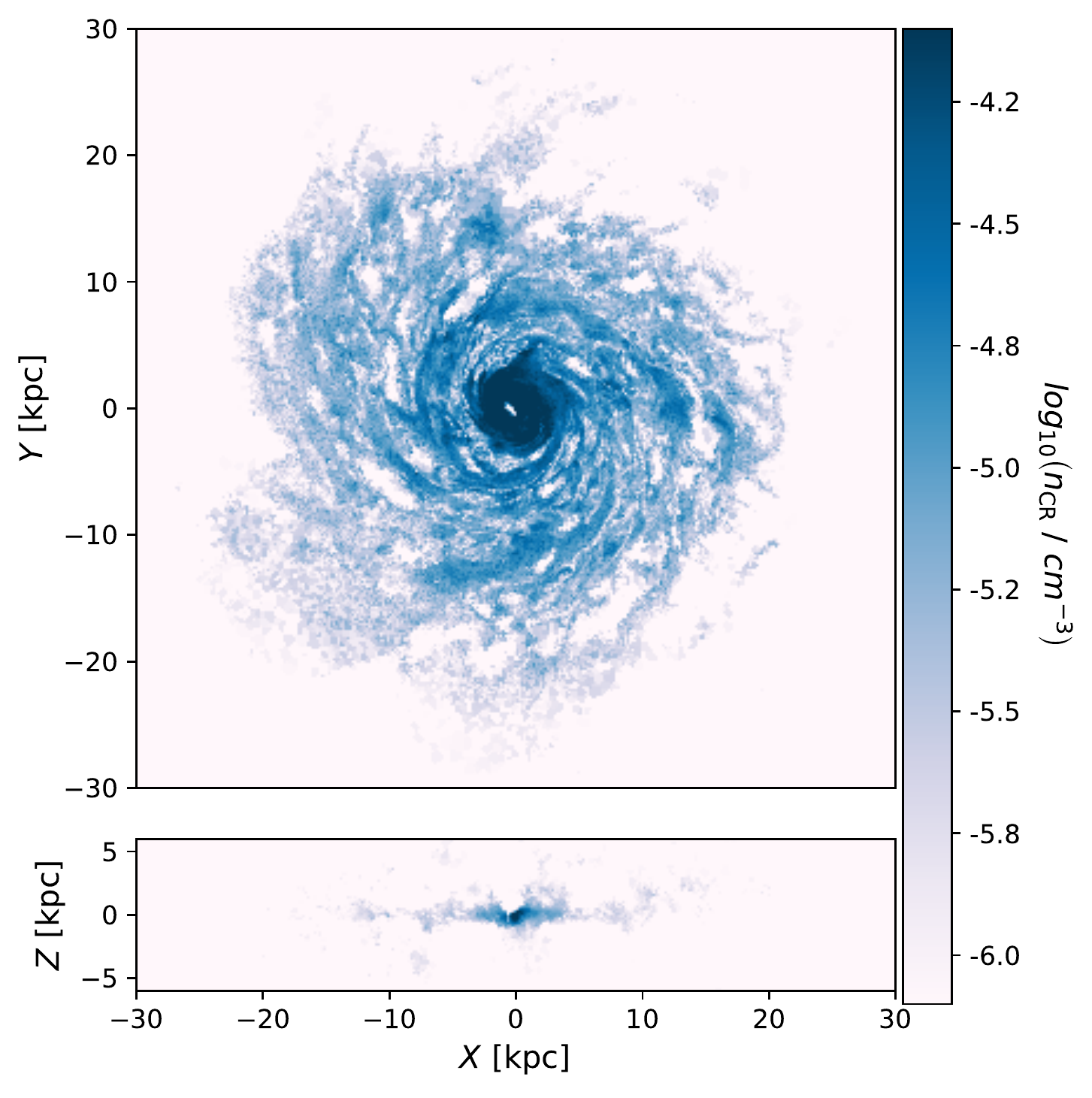}
                                  \vspace{-2mm}
                                  \caption{The same as Figure \ref{fig:MidThB} for CR electron number densities $n_{\mathrm{CR}}$ for the CR1 model (left) and CR2 model (right) derived from Equation \ref{eq:DistCR1} (left panel) and Equation \ref{eq:DistCR2} (right panel), respectively.} 
				  \label{fig:MidCR}
                         \end{center}
                       \end{minipage} 
\end{center} 
\end{figure*}

\subsection{Turbulent magnetic field}
\label{sect:TurbulentB}
While MHD simulations provide the large-scale Galactic magnetic field they usually lack a small scale turbulent component due to insufficient numerical resolution. This component most likely follows a power-law, the exact parameters with regard to scale, magnitude, decay rates needs yet to be constrained \citep[see e.g.][for details]{Rand1989,Minter1996,Han2004,Sun2008}. 

Especially, grid cells covering a regions of several $\mathrm{kpc}$ lack the small scale turbulent component. A more detailed modeling might require a subgrid approach but this is beyond the scope of this paper. Hence, we explore the influence of a turbulent magnetic field component added to our Galactic modeling and code testing setup by means of a Gaussian random field. We note that the original field of the Au-6 simulation already entails some turbulence and numerical noise. However, in the context of this paper we refer to the original Au-6 field as large scale field and to the Gaussian component as turbulent component.

A technique for generating Gaussian fluctuations by means of harmonics of a power-spectrum is presented in \cite{Martel2005}. However, this technique requires a resolution much smaller than Au-6. In order to reproduce the small-scale structures known from all-sky synchrotron emission and FR observations \cite[][]{Haslam1981A,Oppermann2012,Planck2016A} we follow the procedure applied in \cite{Sun2008} and add to the large scale field $\vec{B}_{\rm MHD}(\vec{r})$ a Gaussian random component,
\begin{equation}
\vec{B}(\vec{r})=\vec{B}_{\rm MHD}(\vec{r})+b_{\rm \sigma_b}\vec{N}_{\rm \sigma_B}\, ,
\end{equation}
Here, $\vec{N}_{\rm \sigma_B}$ is the normalised direction with an angle between $\vec{N}_{\rm \sigma_B}$ and $\vec{B}_{\rm MHD}$ randomly sampled from a Gaussian with a variance of $\sigma_B=25^{\circ}$ selected to resemble observed small scale structures (see also Section \ref{sect:ObsSynchrotron}). The magnitude of the Galactic turbulent field is estimated to be $\approx 2-3\ \rm{\mu G}$ \citep[][]{Sun2008}. Thus, we sample from a Gaussian with $\sigma_b=2\ \rm{\mu G}$. We set $b_{\rm \sigma_b}=|\vec{B}_{\rm MHD}|$ in case of $b_{\rm \sigma_b}>|\vec{B}_{\rm MHD}|$. This is also in agreement with the finding of \cite{Han2004,Han2006}. They report that the Galactic regular field is of the same order of magnitude as the turbulent component. Here, we do not attempt to keep the field divergent free since this is not of relevance to RT simulation as outlined in the following sections.

\begin{figure*}
 \begin{center}
         \begin{minipage}[c]{1.0\linewidth}
                         \begin{center}
                                 \includegraphics[width=0.49\textwidth]{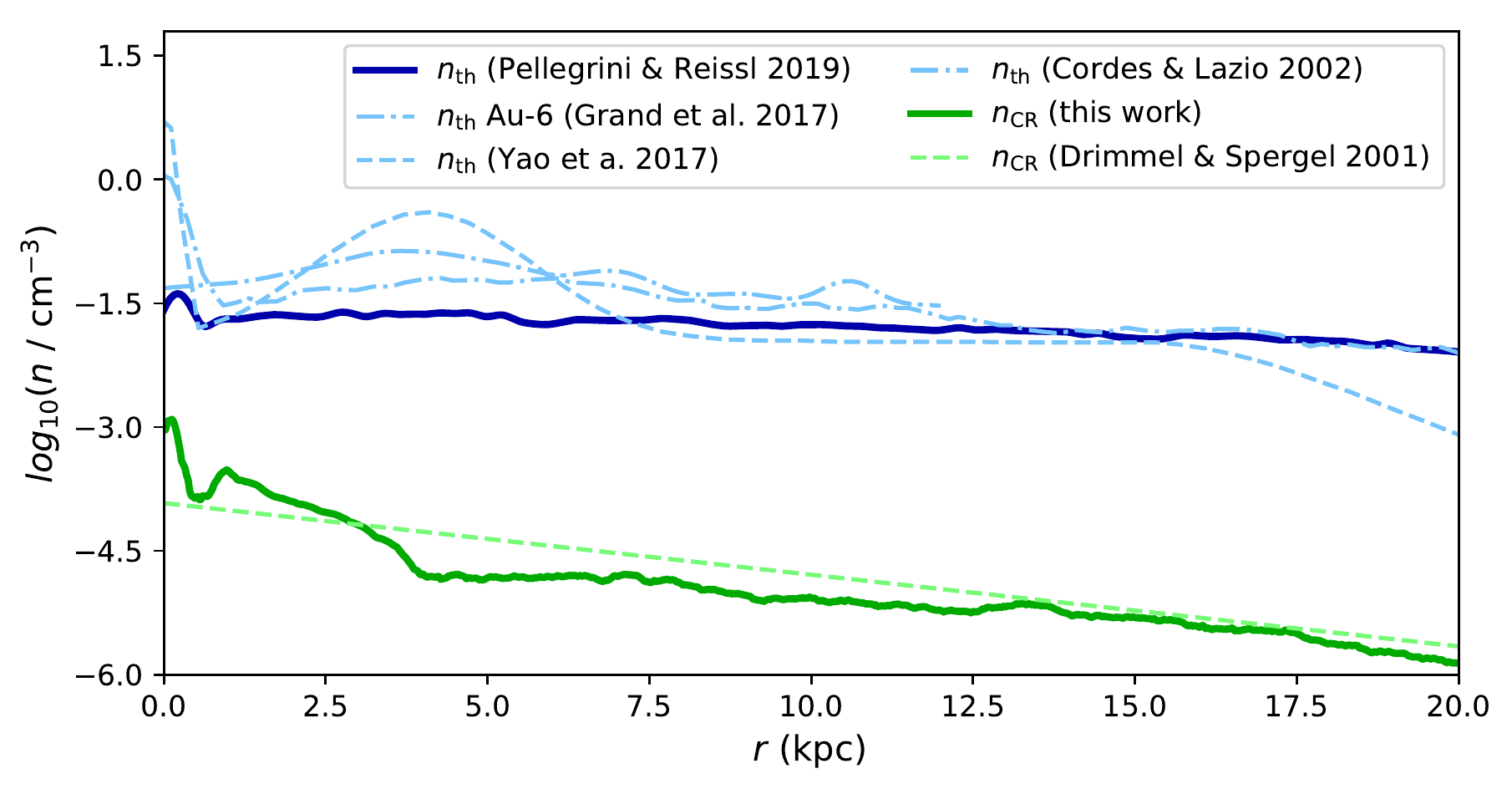}
                                 \includegraphics[width=0.49\textwidth]{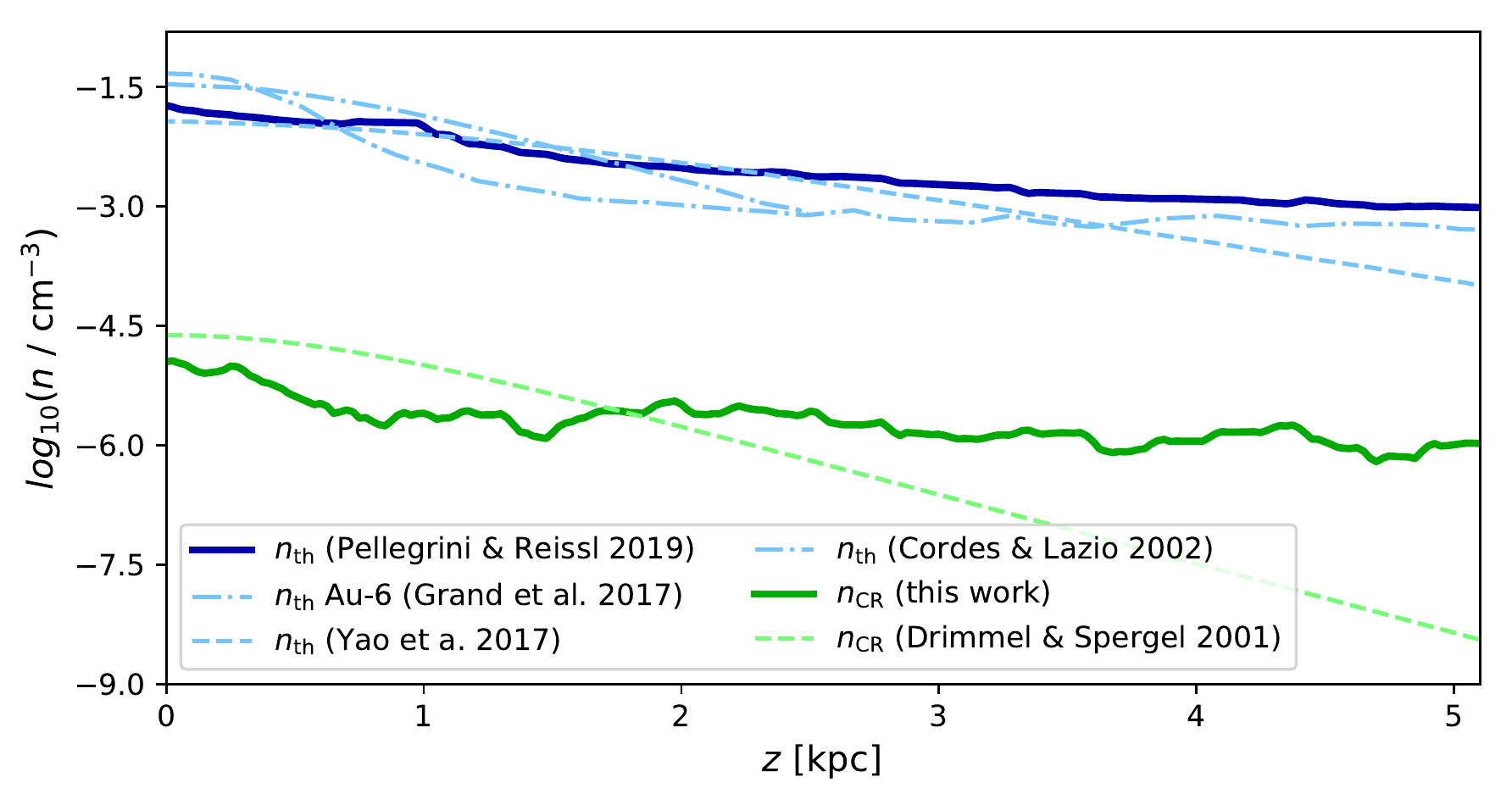}
                                 
                                  \caption{Left panel: Comparison of thermal electron number density $n_{\mathrm{th}}$ resulting from the population synthesis model of \protect\cite{ReisslA} with the estimates presented in \protect\cite{Yao2017} and \protect\cite{Cordes2002} (blue lines) and a comparison of the parametrized CR1 electron density $n_{\mathrm{CR}}$ presented in \protect\cite{Drimmel2001} with the CR2 mode derived in Section \ref{sect:CRDistribution} (green lines). All quantities are averaged along the Galactic radius $r$ in the disc midplane $z=0\ \mathrm{kpc}$. Right panel: The same, but averaged along the $z$-direction at the solar radius $r=8\ \mathrm{kpc}$.}
                                  
                                  \label{fig:ProfDens}
                         \end{center}
                       \end{minipage} 
											
			         \begin{minipage}[c]{1.0\linewidth}
                         \begin{center}
                                 \includegraphics[width=0.49\textwidth]{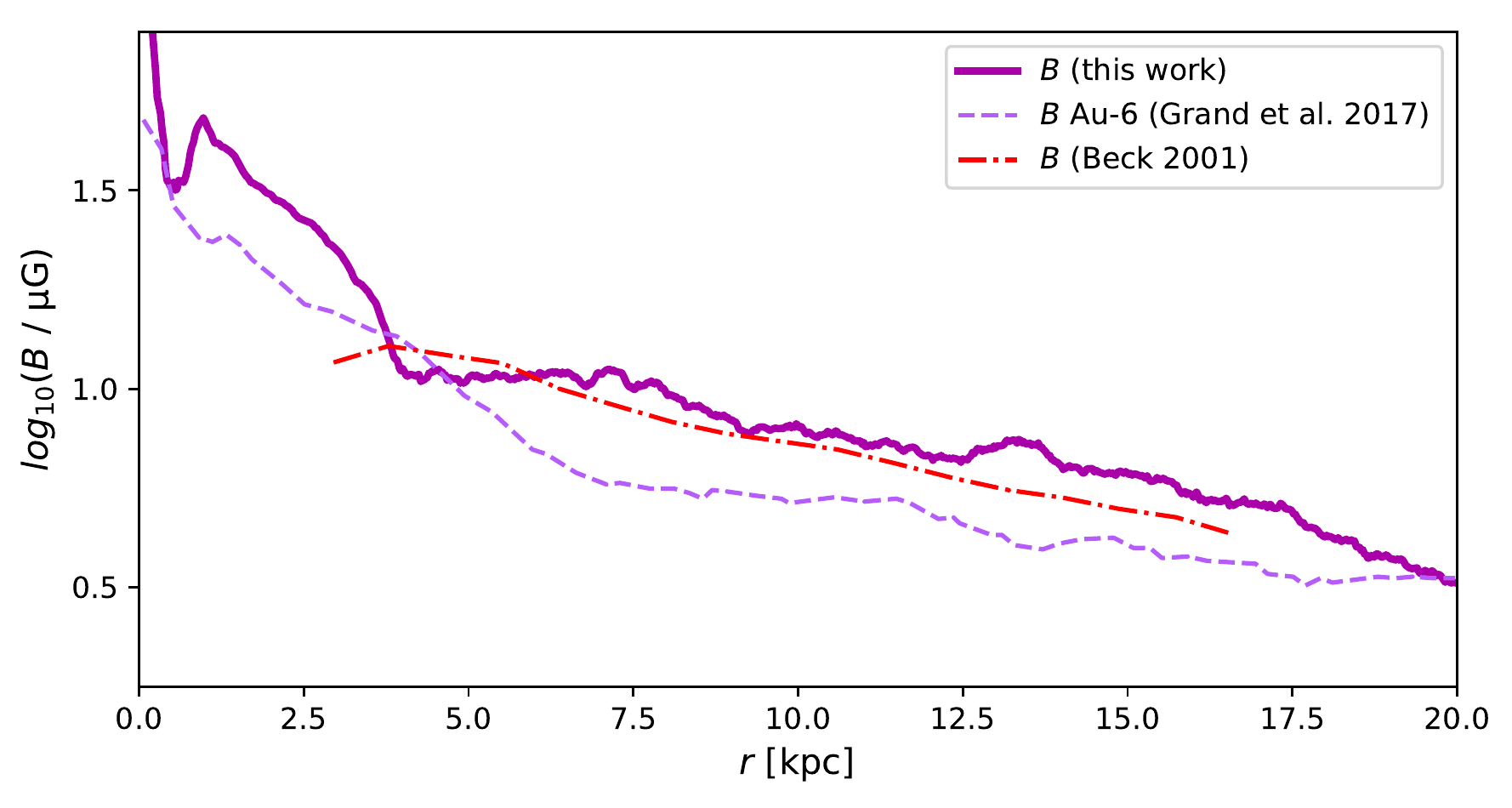}
                                 \includegraphics[width=0.49\textwidth]{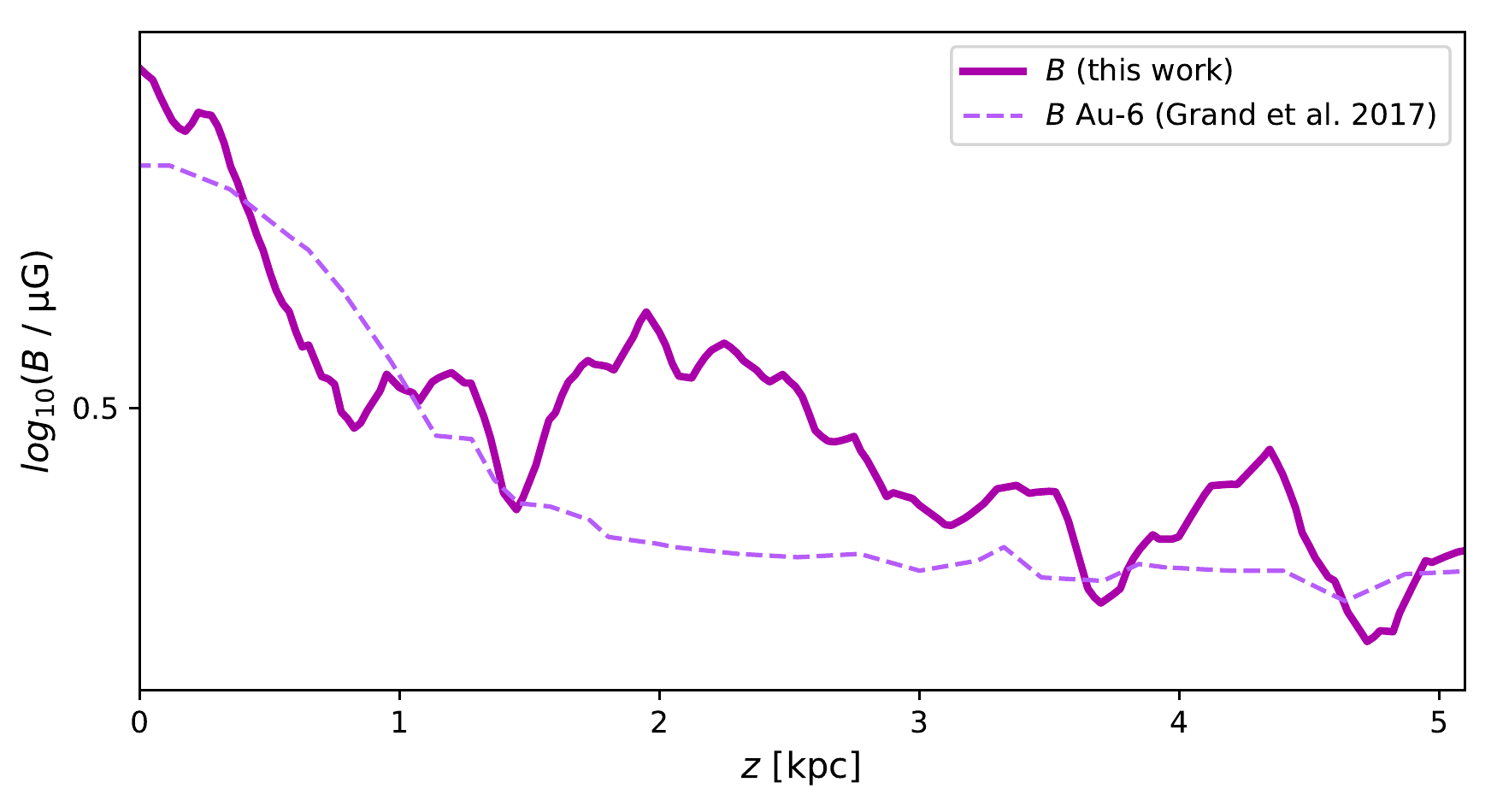}
                                  \caption{The same as Figure \ref{fig:ProfDens} for the magnetic field strength $B$ derived in Section \ref{sect:TurbulentB} in comparison with the original Au-6 field strength \citep[see][]{Grand2017} and with the data presented in \protect\cite{Beck2001}.}
                                  
                                  \label{fig:ProfB}
                         \end{center}
                       \end{minipage} 											
\end{center} 
\end{figure*}
\subsection{CR electron distribution}
\label{sect:CRDistribution}

Cosmic ray electrons do not seem to be closely connected to the overall gas density structures of the galaxy \citep[][]{Strong2004,Page2007}. A smooth parameterisation of the CR electron distribution is suggested in \cite{Drimmel2001} and \cite{Page2007}, respectively, as
\begin{equation}
n_{\rm CR_1}(x,y,z)=n_{\rm 0,CR}\exp\left( -\frac{r}{h_{\rm r}} \right) \cosh^{-2} \left(\frac{|z|}{h_{\rm z}} \right)\, .
\label{eq:DistCR1}
\end{equation}
Here, $x$, $y$, $z$ are Cartesian coordinates with their origin at the Galactic centre in units of $\mathrm{kpc}$ and $r=\sqrt{x^2+y^2}$ is the galactocentric radius. We apply a characteristic radius $h_{\rm r}=5\ \rm{kpc}$ and a scale height of $h_{\rm z}=1\ \rm{kpc}$ \citep[][]{Sun2008} for the CR distribution. The central CR electron density is taken to be the $n_{\rm 0,CR}=1.74\ 10^{-4} \rm{cm^{-3}}$ selected such that we ensure a  typical value of $n_{\rm Earth}=6.4\ 10^{-5} \rm{cm^{-3}}$ for our solar neighbourhood at $r = 8\ \rm{kpc}$ and $z = 0\ \rm{kpc}$. Although, it remains to be seen if this density is typical for the entire Milky-Way. Here, we refer to this parametrization as the CR1 model.

Alternatively, the CR electron distribution can be derived from an equipartition argument. The magnetic energy stored in a cell of a certain volume is given by
\begin{equation}
u_{\rm B}=\frac{1}{8\pi}B^2\, .
\end{equation}
The total energy density of the CR electrons can be estimated by integrating the particle energy $\gamma m_{\rm e} c^2$ over the CR distribution function. This results in
\begin{equation}
\begin{split}
u_{\rm CR}=\int_{\gamma_{\rm min}}^{\gamma_{\rm max}} {\gamma m_{\rm e} c^2}N_{\rm CR}(\gamma)d\gamma \approx  \qquad\qquad\qquad\qquad \\ \qquad\qquad\qquad\qquad n_{\rm CR}  m_{\rm e} c^2 \gamma_{\rm min}\frac{p-1}{p-2}
\end{split}
\end{equation}
given that $p\neq 2$ and $\gamma_{\rm min} \ll \gamma_{\rm max}$. We assume the magnetic field to be in equipartition with the CR electrons ($u_{\rm B}=u_{\rm CR}$). Consequently, the CR distribution function (Equation \ref{eq:CRDistribution}) scales approximately with the local magnetic field $B$ via:
\begin{equation}
 n_{\rm CR_2}(x,y,z) \simeq \frac{B^2(x,y,z) (p-2)}{8\pi \gamma_{\rm min}(p-1)m_{\rm e}c^2}
 \label{eq:DistCR2}
\end{equation}
We refer to this relation as the CR2 model as an alternative synchrotron RT test scenario for {\sc POLARIS}. 

The power law index $p$ as well as the lower cut off $\gamma_{\rm min}$ and upper cut off $\gamma_{\rm max}$ of the distribution function  (see Equation \ref{eq:CRDistribution}) are also not well constrained galaxy-wide. The range between the cut offs is usually taken to be between the order of unity and several hundred \citep[e.g.][]{Ferland1984,deKool1989,Strong2011}. In this work we apply a typical value of $\gamma_{\rm min} = 4$ \citep[see e.g][]{Webber1998} and $\gamma_{\rm max} = 300$, respectively, for the CR spectrum. Indeed, the exact value of the upper cut off is of minor relevance as long as $\gamma_{\rm max} \gg 1$ (compare Equation \ref{eq:CRJI} and Equation \ref{eq:CRAlphaI}). 

In principle the power-law index $p$ may change dependent on the position within the galaxy. Observations suggested that $p$ varies with height $z$ \citep[e.g.][]{Miville2008} with lower values in the plane and large values toward the halo with a range of about $2-4$ \citep[][]{Bennett2003,Miville2008}.

For our galaxy models we assume a fixed power-law index with a canonical value of $p=3$ \citep[][]{Rybicki1979,Miville2008}. Consequently, the maximal degree of linear polarisation that can be expected from synchrotron emission is $P_{\rm l}=0.75$ (see Equation \ref{eq:MaxPl}).

\section{Results and Discussion}
\label{sect:ResultsDiscussion}

With the above new features and extensions of {\sc POLARIS} applied to the Au-6 galaxy, we now discuss in detail how the results compare to the observed characteristics of the Milky Way and of selected extragalactic systems. 
\begin{figure*}
 \begin{center}
         \begin{minipage}[c]{1.0\linewidth}
                         \begin{center}
                                 \includegraphics[width=0.49\textwidth]{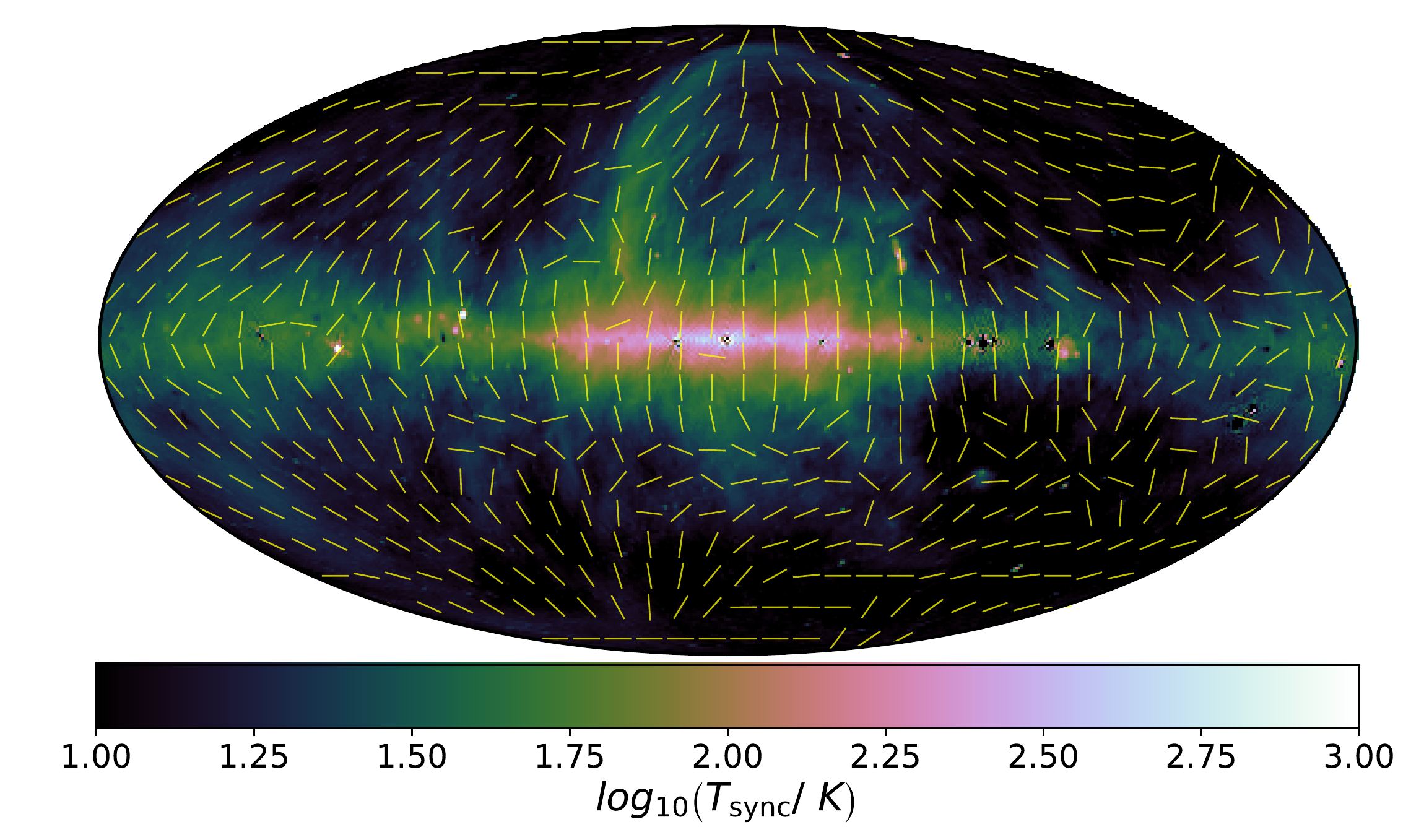}
                                 \includegraphics[width=0.49\textwidth]{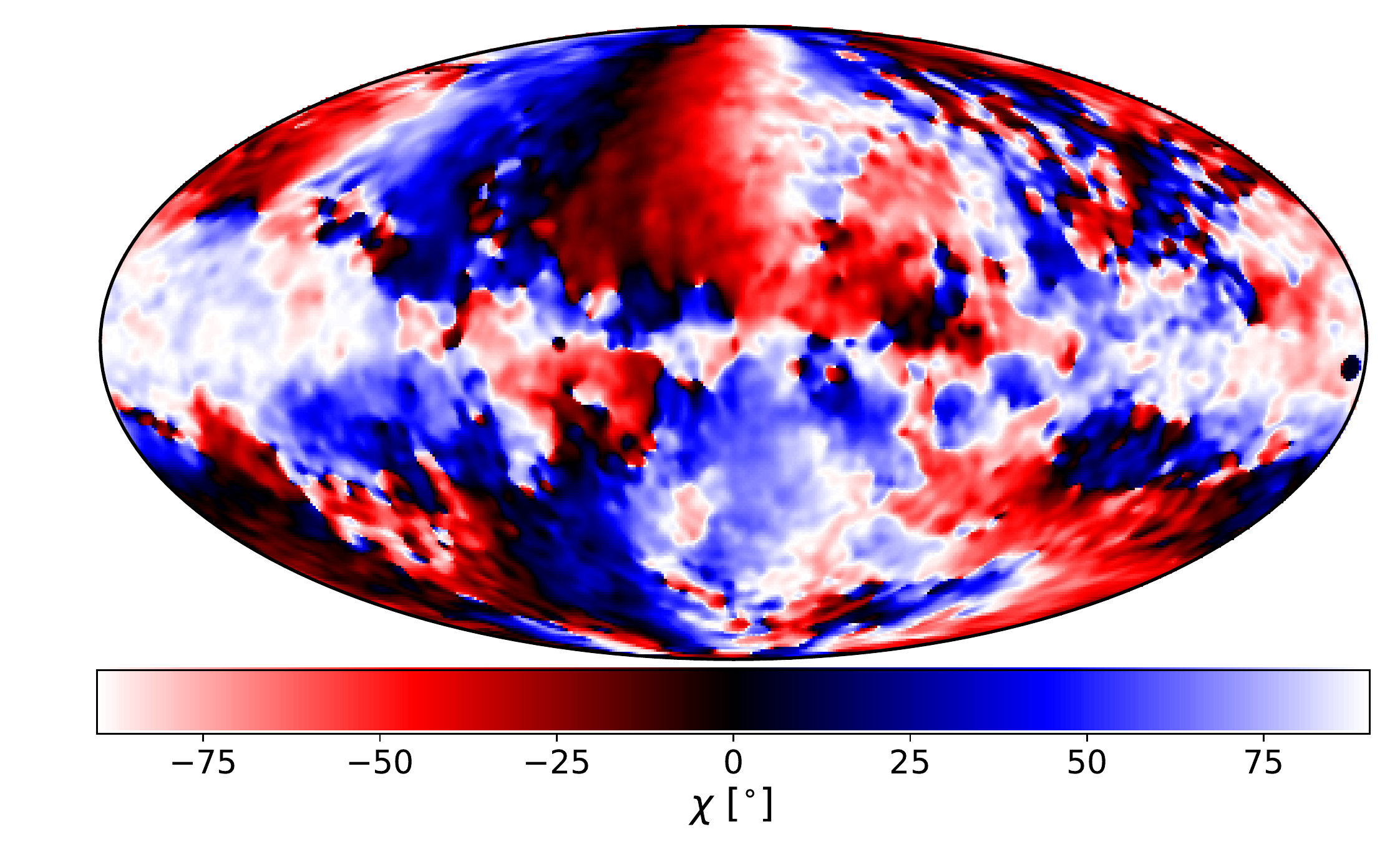}
                         \end{center}
                       \end{minipage} 
         \begin{minipage}[c]{1.0\linewidth}
                         \begin{center}
                                 \includegraphics[width=0.49\textwidth]{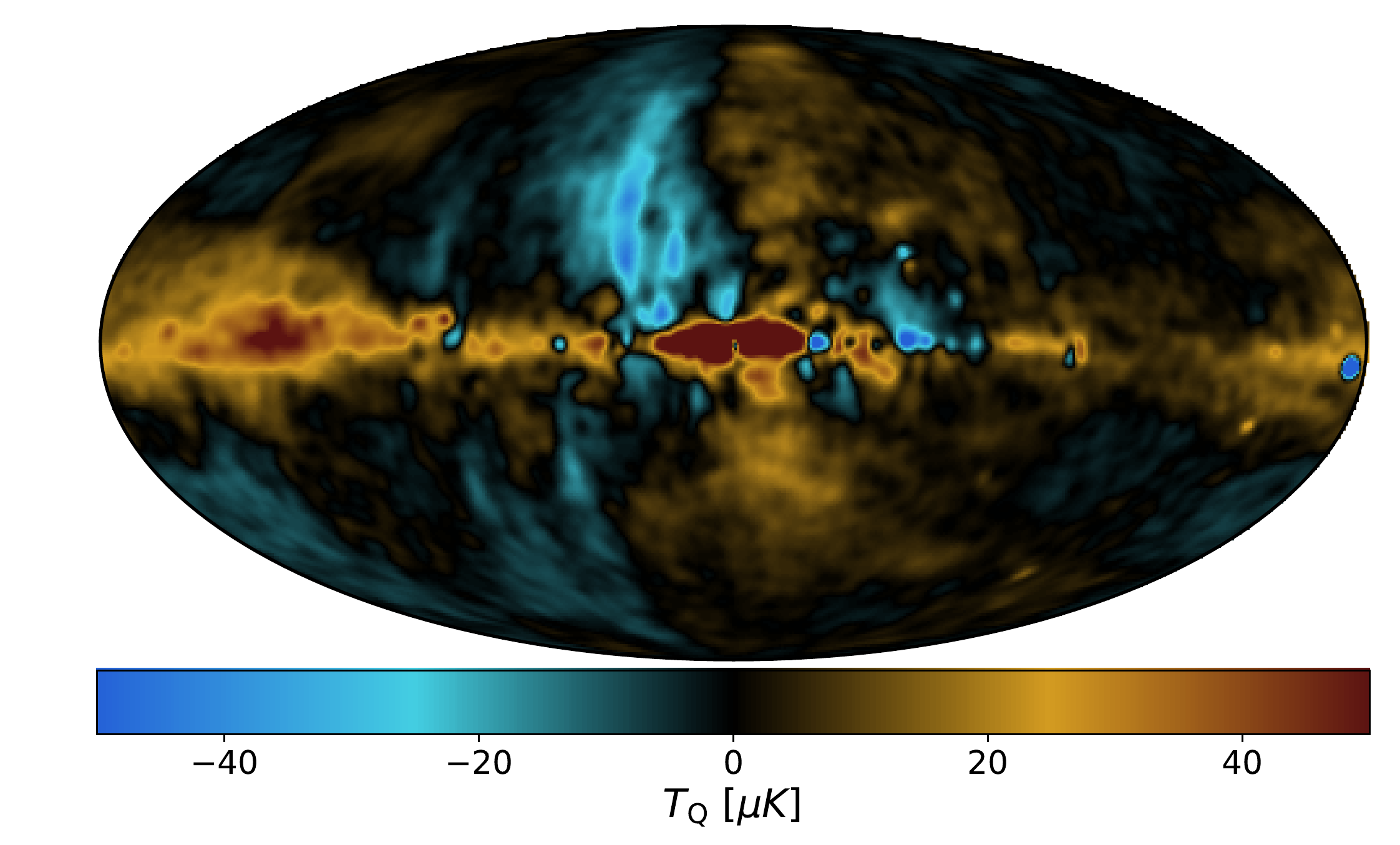}
                                 \includegraphics[width=0.49\textwidth]{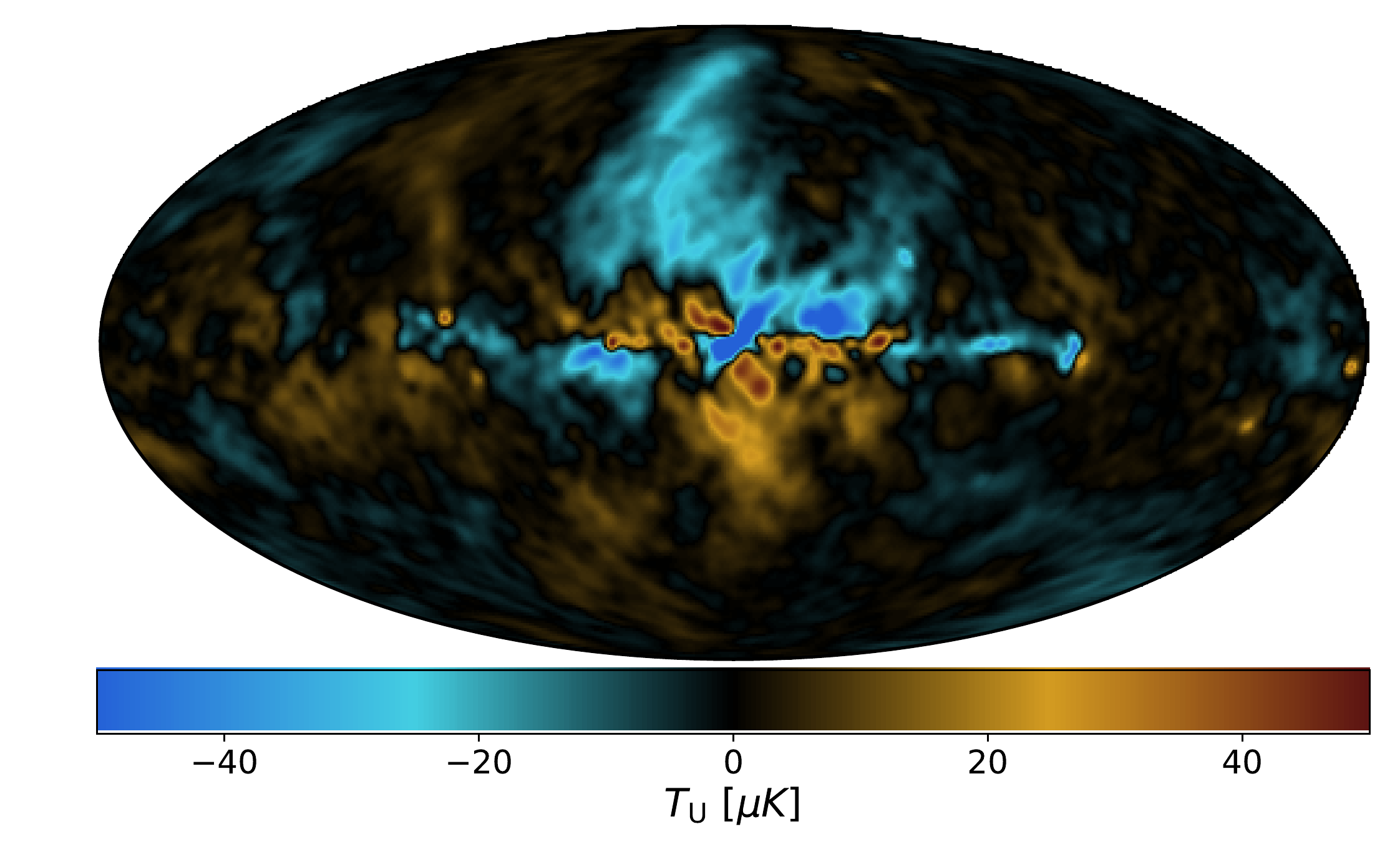}
                         \end{center}
                       \end{minipage} \end{center}  
 \caption{Top Row: Synchrotron emission at $734.8\ \mathrm{mm}$ (left panel) overlaid with polarisation vectors derived from the $1.0\ \mathrm{mm}$ polarisation observations \citep[see][for details]{Haslam1981A,Haslam1982B} and the orientation angles of the Galactic synchrotron polarisation (right panel) at $1.0\ \mathrm{mm}$ (see Equation \ref{eq:LinearOrientation}). Bottom row: Corresponding $1.0\ \mathrm{mm}$ synchrotron Stokes Q map (left panel) and Stokes U map (right panel) \citep[see][]{Page2007,Hinshaw2009}.}
 \label{fig:AllSkyPlanck}
\end{figure*}
\subsection{Galactic observables}
\label{sect:ResultsProfiles}
The spatial distribution of the key input parameters necessary for computing realistic synchrotron emission maps is illustrated in Figures \ref{fig:MidThB} and \ref{fig:MidCR}. We show cuts through the disc midplane and perpendicular to it. The thermal electron number density is presented in the left panel of Figure \ref{fig:MidThB} based on the population synthesis model presented in \cite{ReisslA}, while the right panel gives the magnetic field structure of the Au-6 galaxy \citep[][]{Grand2017} extended by a turbulent component, as discussed in Section \ref{sect:TurbulentB}. Magnetic field strength and thermal electrons clearly correlate and exhibit a characteristic spiral structure, as expected from extragalactic observations \citep[e.g.][]{Beck2001}. The distribution of cosmic ray electrons, CR1 and CR2, as discussed in the previous section, is shown in Figure \ref{fig:MidCR}. Comparing both distributions one might expect more small-scale features in the synthetic synchrotron observations from model CR2.

A more quantitative view of these distributions is given in Figures \ref{fig:ProfDens} and  \ref{fig:ProfB}, where we present the average radial profile in the disc midplane as well as the vertical profile out of the disc computed at the solar radius of $8\ \mathrm{kpc}$. The thermal electron distribution of the population synthesis model of \cite{ReisslA} (blue line) agrees well with the data of \cite{Yao2017} as well as with the the model of \cite{Cordes2002},  with only small deviations close to the galactic centre and the very outer disc. As similar behavior is visible in the distribution of cosmic ray electrons (green line), which we compare to the parameterisation of \protect\cite{Drimmel2001}. In the range $2.5\ \mathrm{kpc}<r<17\ \mathrm{kpc}$ and $z<3\ \mathrm{kpc}$ our model is able to reproduce the existing data and other theoretical models very well in terms of the overall density of free electrons. We note that deviations in the galactic centre and in the outer parts of the disc contribute very little to the observed emission on the sky of an observer at the solar neighbourhood at $r\approx 8\ \mathrm{kpc}$ and $z\approx 0\ \mathrm{kpc}$ \citep[][]{Pakmor2018}. Similar holds for the strengths of the magnetic field which enters our calculation of synchrotron emission and Faraday rotation. Our model very well reproduces the data presented by \cite{Beck2001} in the disc midplane. However,  our field is typically 10-20\% stronger than the original magnetic field of the Au-6 galaxy \citep[see][]{Grand2017} because we add a turbulent component as introduced in Section \ref{sect:TurbulentB}.

\begin{figure*}
 \begin{center}
         \begin{minipage}[c]{1.0\linewidth}
                         \begin{center}
                                 \includegraphics[width=0.49\textwidth]{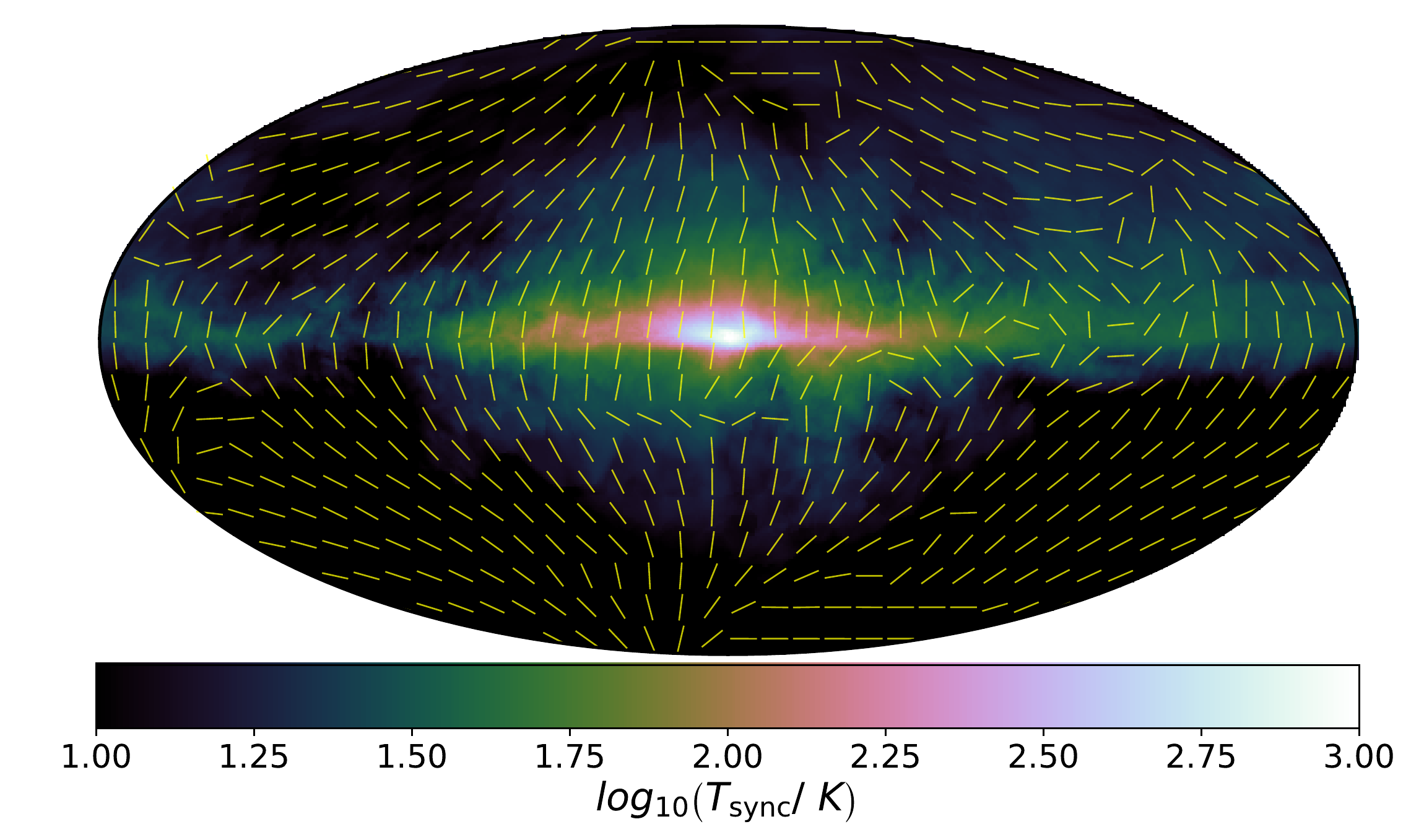}
                                 \includegraphics[width=0.49\textwidth]{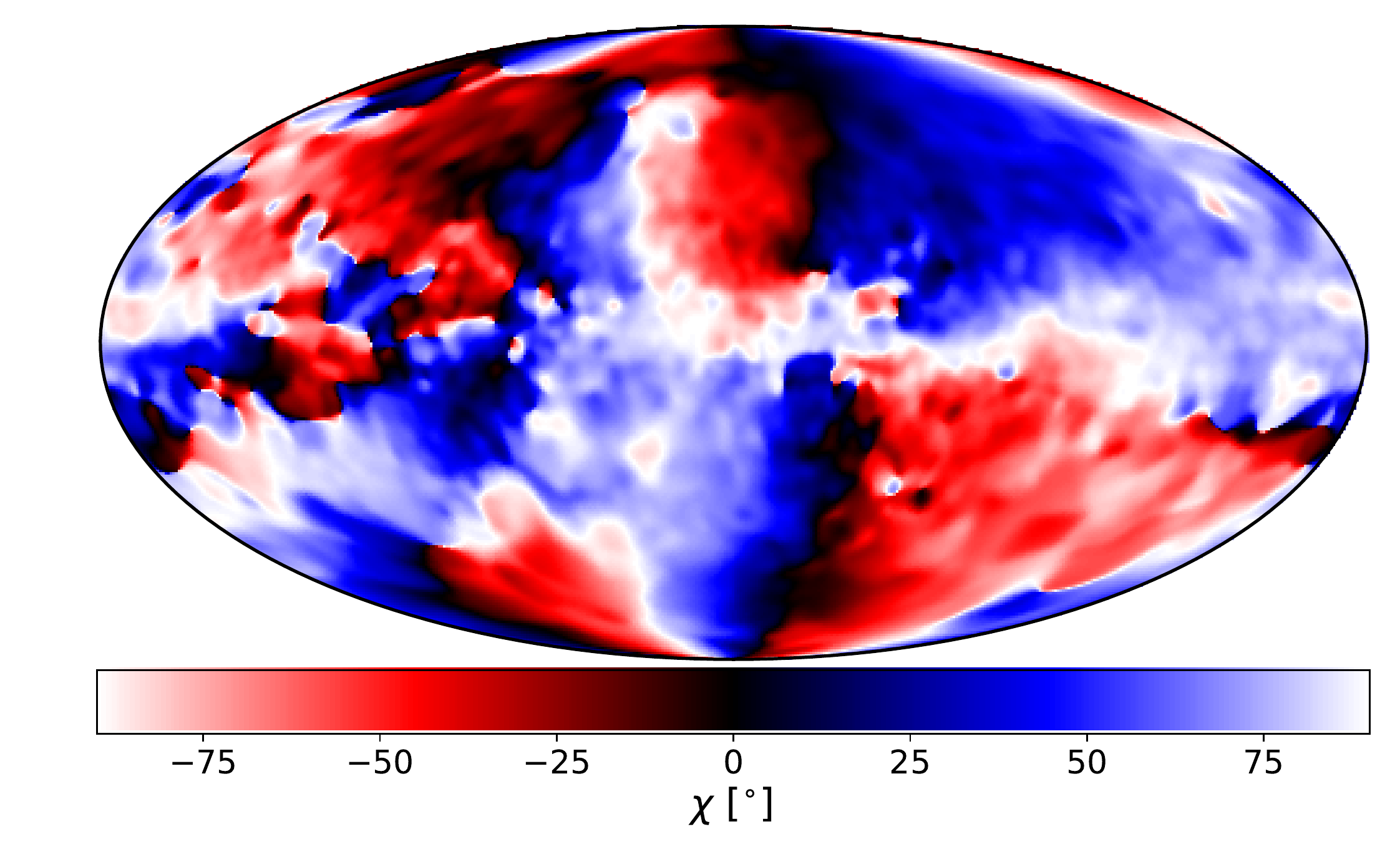}
                         \end{center}
                       \end{minipage} 
         \begin{minipage}[c]{1.0\linewidth}
                         \begin{center}
                                 \includegraphics[width=0.49\textwidth]{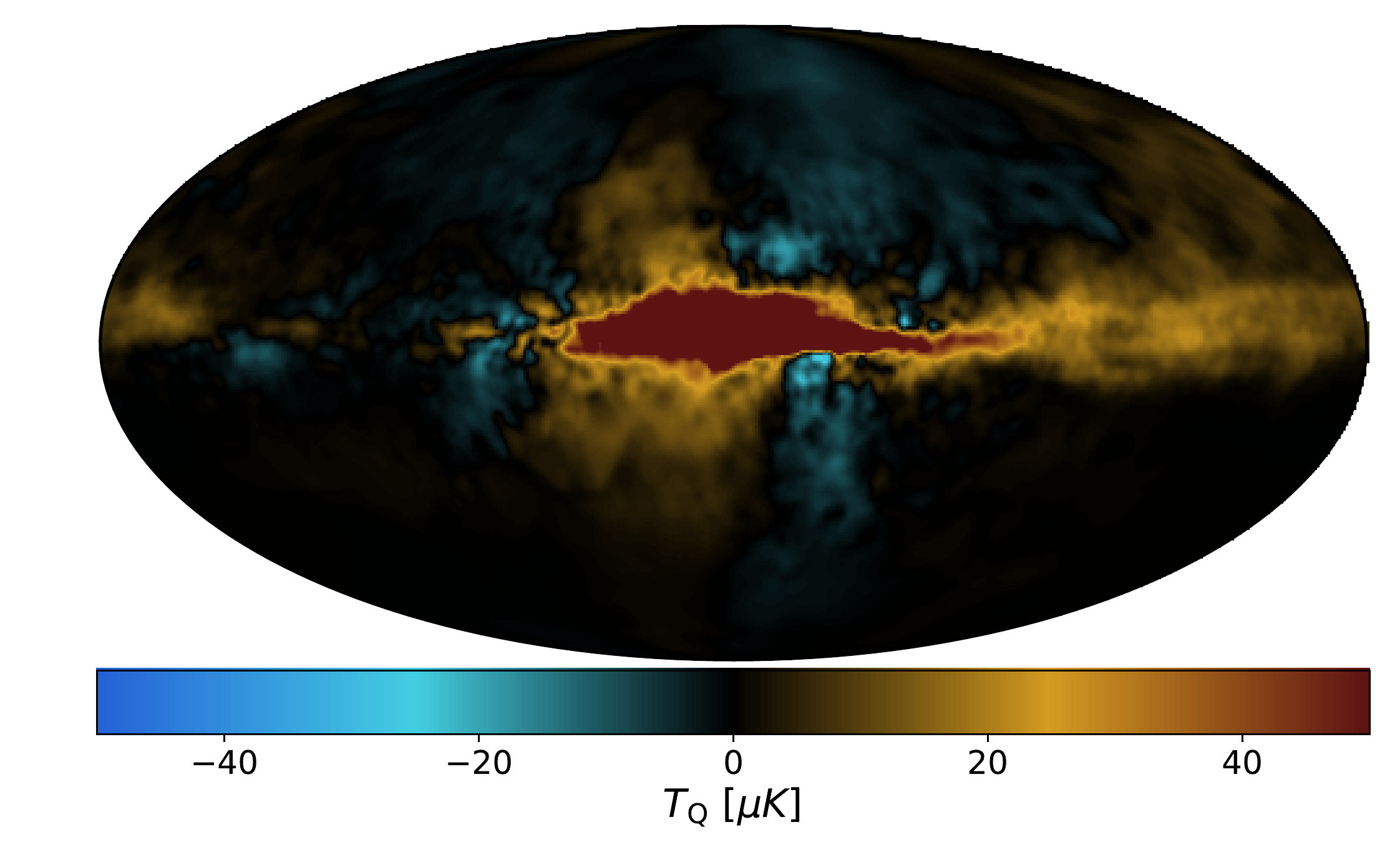}
                                 \includegraphics[width=0.49\textwidth]{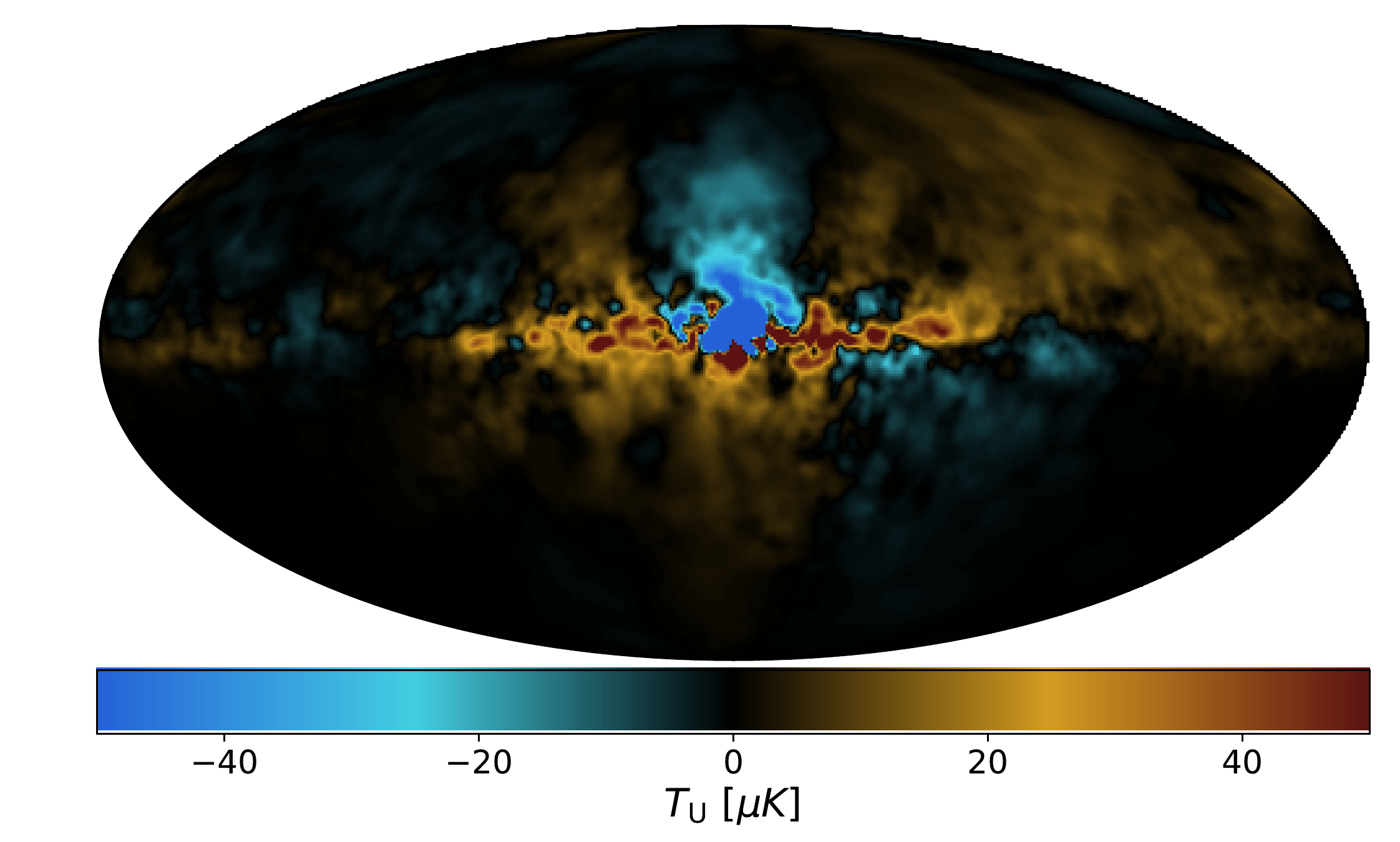}
                         \end{center}
                       \end{minipage} \end{center}  
 \caption{The same as Figure \ref{fig:AllSkyPlanck} but now for the synthetic emission resulting for the CR1 electron distribution (see Equation \ref{eq:DistCR1}) as seen from the exemplary observes position P01.}
 \label{fig:AllSkyCR1}
\end{figure*}

\begin{figure*}
 \begin{center}
         \begin{minipage}[c]{1.0\linewidth}
                         \begin{center}
                                 \includegraphics[width=0.49\textwidth]{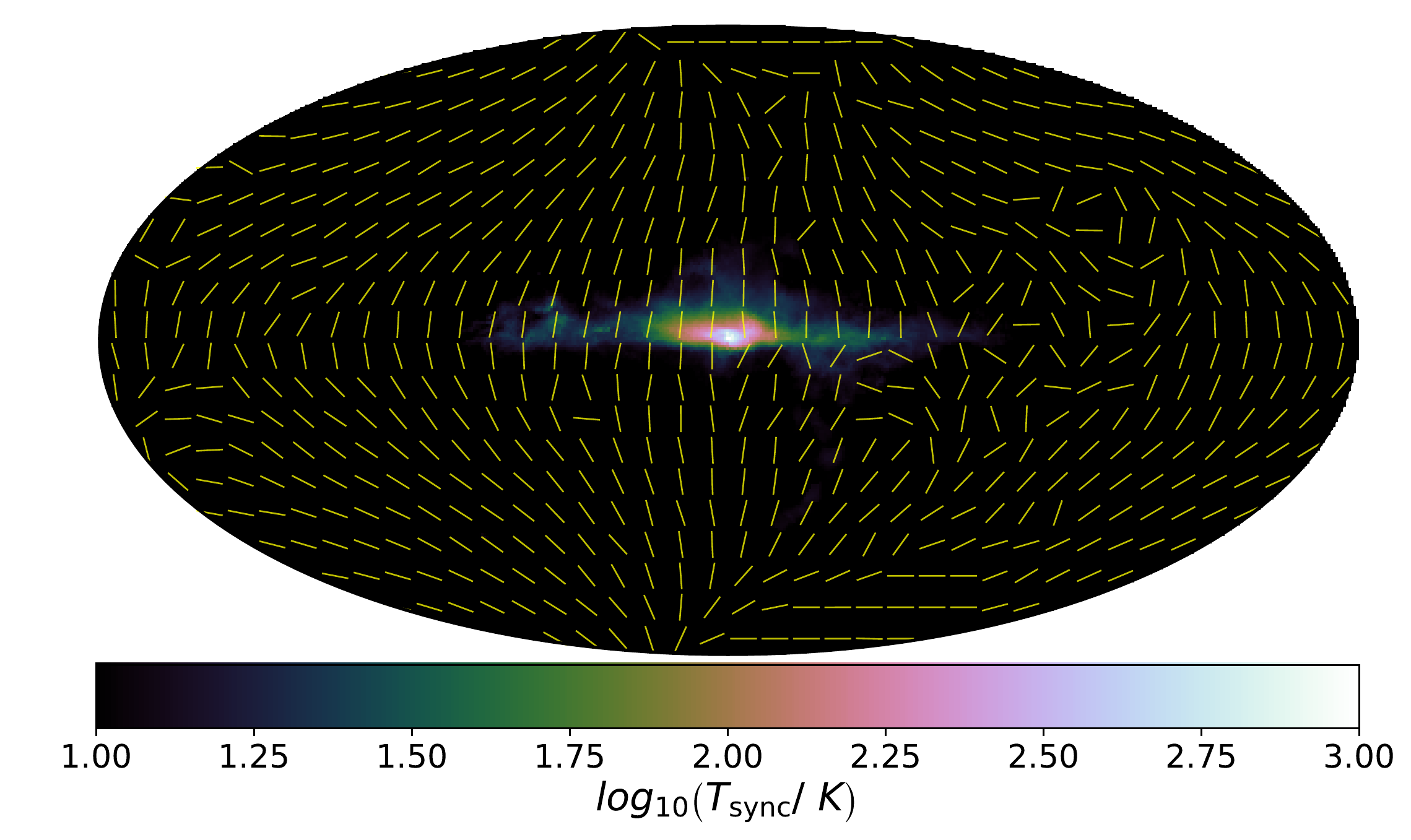}
                                 \includegraphics[width=0.49\textwidth]{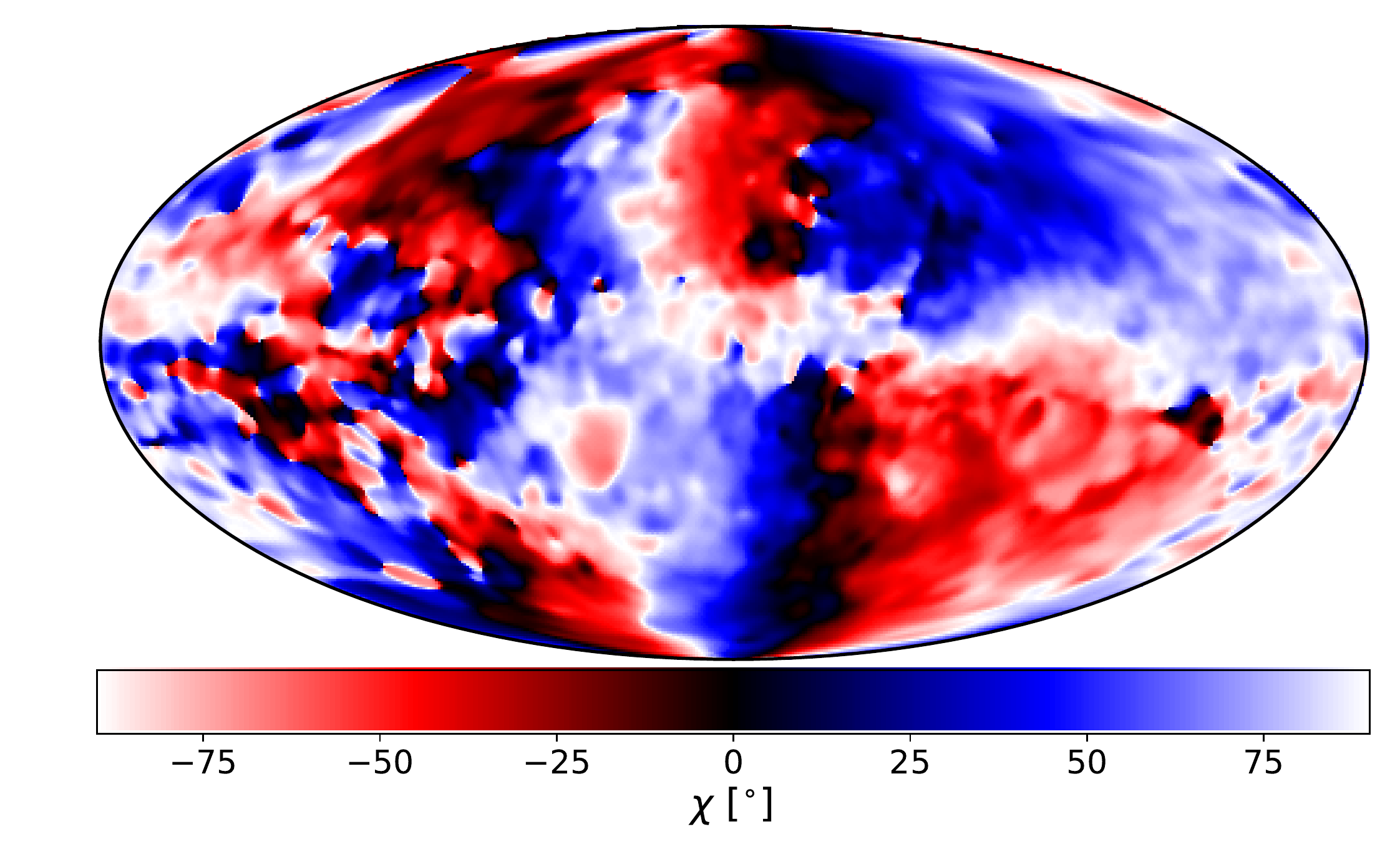}
                         \end{center}
                       \end{minipage} 
         \begin{minipage}[c]{1.0\linewidth}
                         \begin{center}
                                 \includegraphics[width=0.49\textwidth]{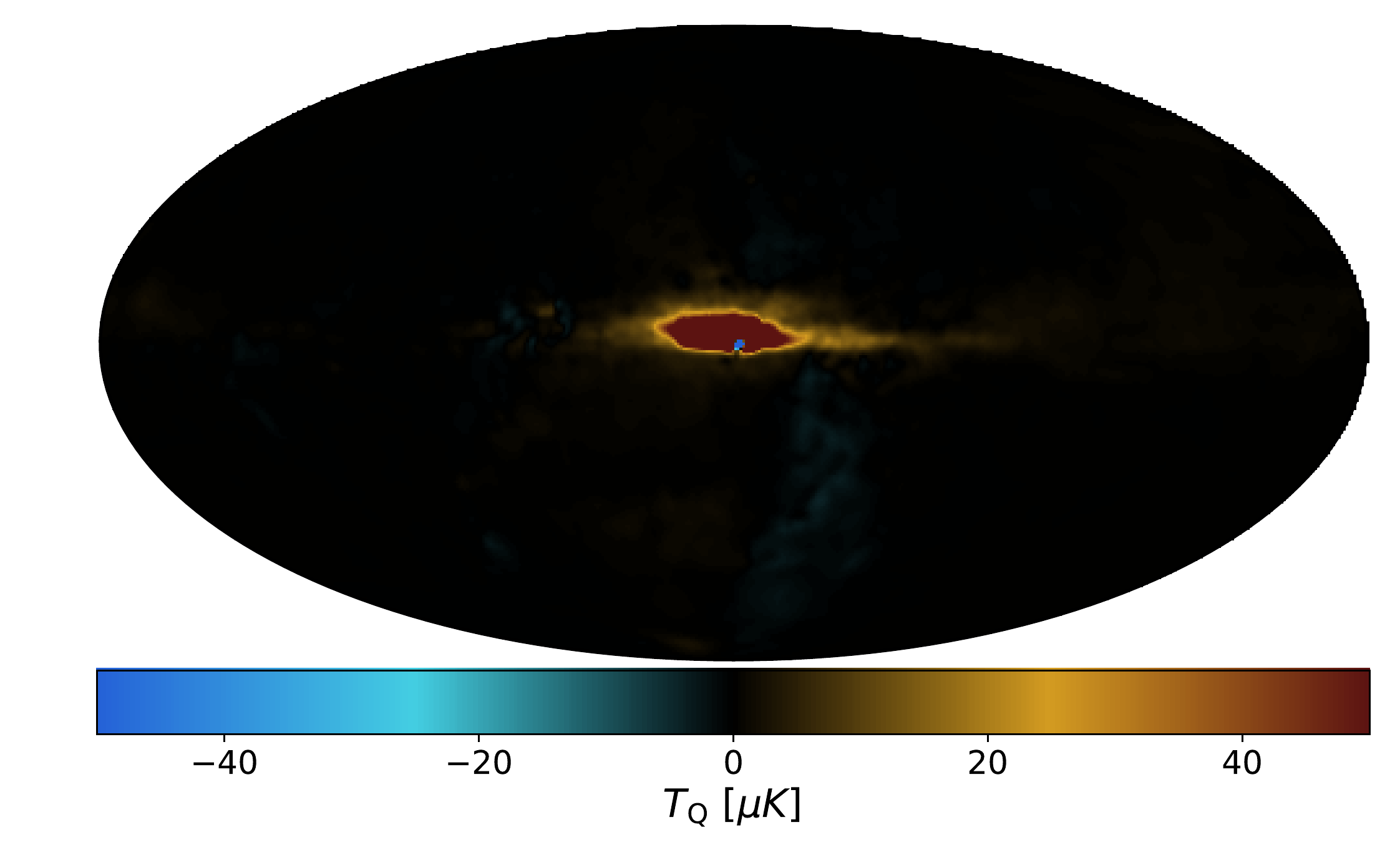}
                                 \includegraphics[width=0.49\textwidth]{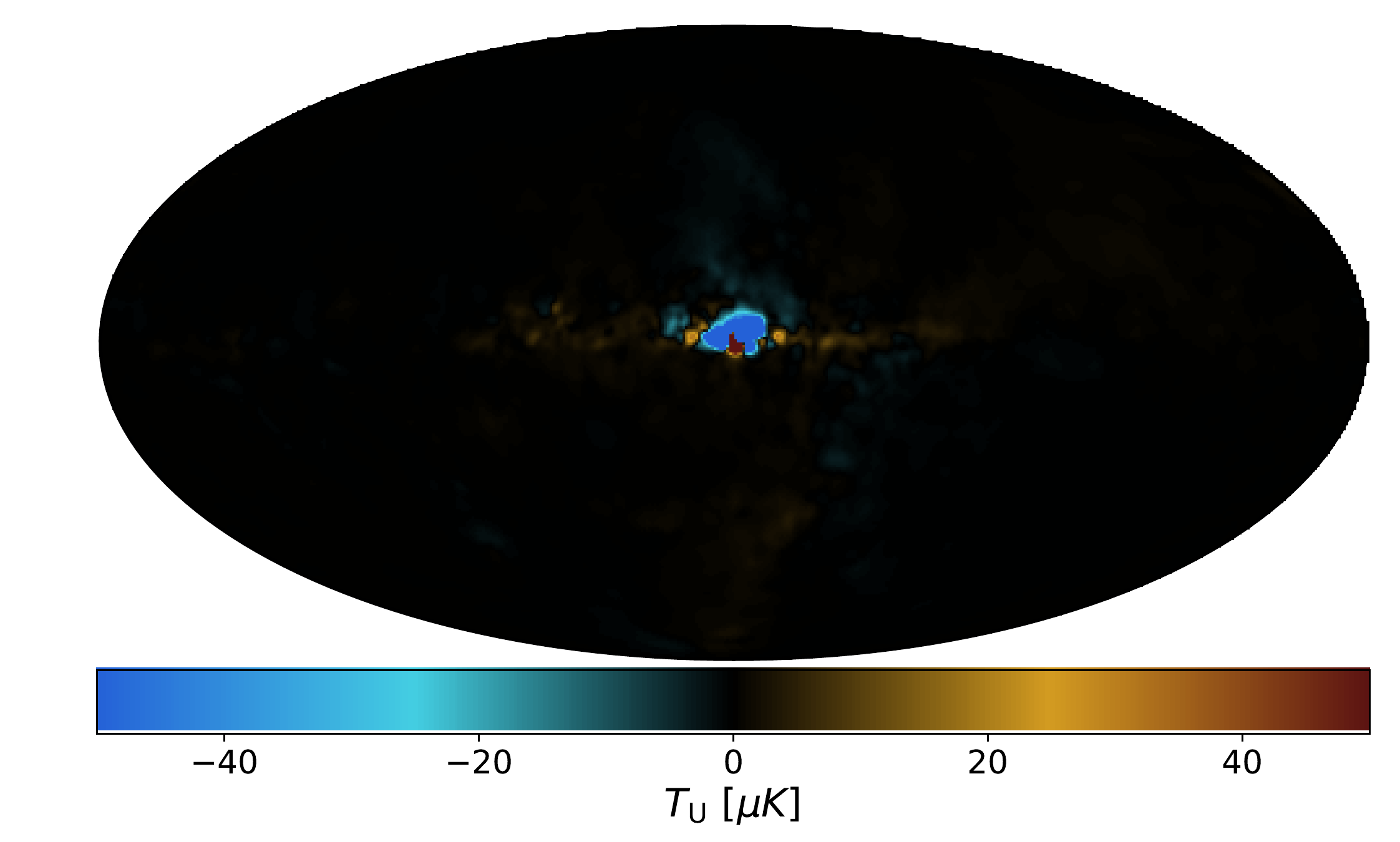}
                         \end{center}
                       \end{minipage} \end{center}  
 \caption{The same as Figure \ref{fig:AllSkyCR1} for the CR2 electron distribution (see Equation \ref{eq:DistCR2}) as seen from the exemplary observes position P01.}
 \label{fig:AllSkyCR2}
\end{figure*}

\begin{figure}
	\begin{center}
			\begin{minipage}[c]{1.0\linewidth}
				\begin{center}
					\includegraphics[width=1.0\textwidth]{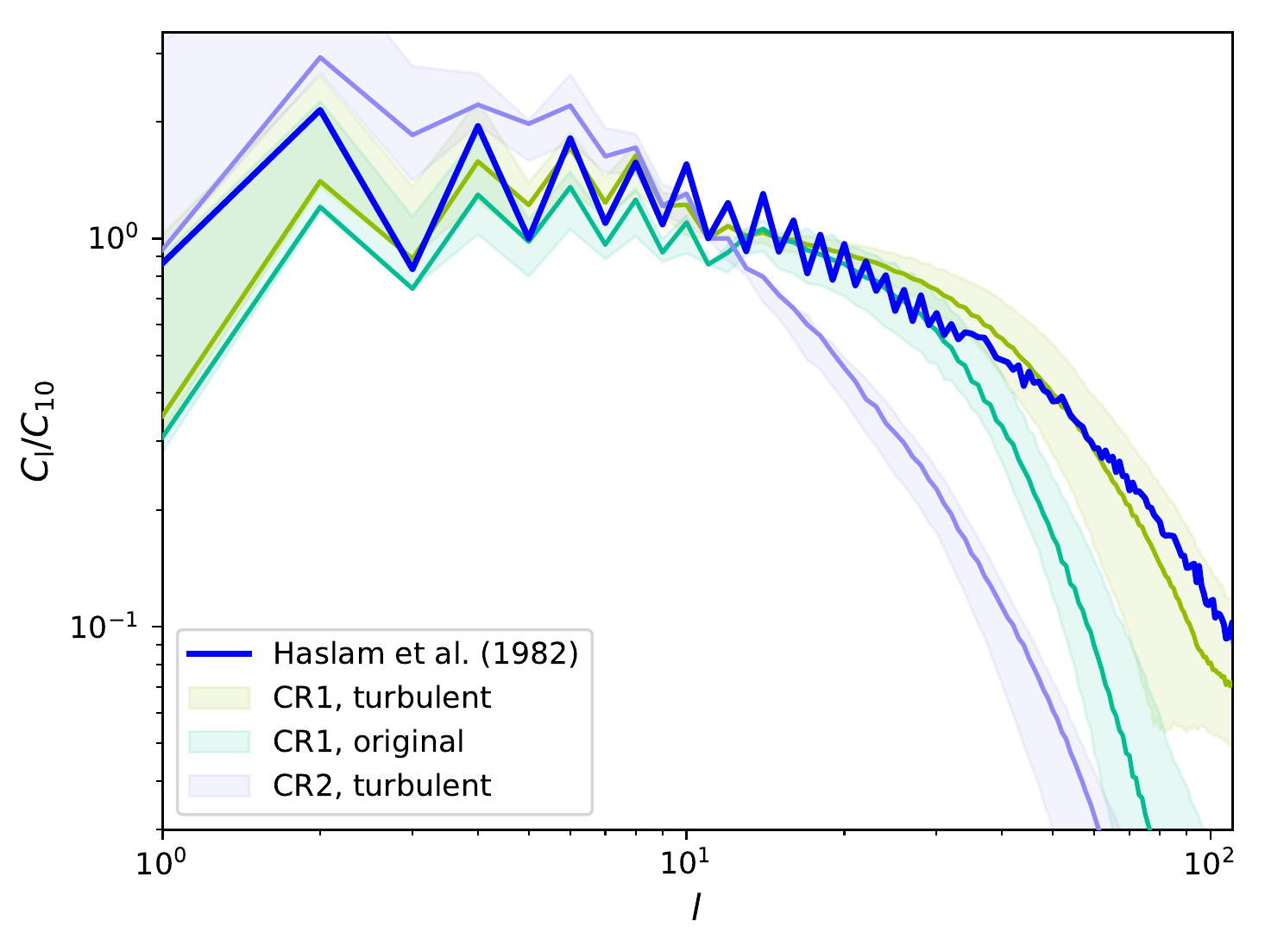}
			\end{center} 
			
			\caption{{Multipole spectrum as a function of multipole moment $l$ of the synthetic all-sky synchrotron emission maps for the distinct observer positions resulting from the CR1 model with turbulent magnetic field component (yellow) CR1 model with the original Au-6 field (green) and the CR2 model with the turbulent field (purple).  For comparison we plot the spectrum of the \protect\cite{Haslam1982B}  map in blue. All spectra are normalised at $l=10$ for a better comparison. Solid lines are for the observer position P01 while the bands are the range of the spectra of the positions P02-P01.}}
			\label{fig:SpectrumSync}
	\end{minipage}
\end{center}
\end{figure}

\subsection{Synchrotron emission and polarisation observations}
\label{sect:ObsSynchrotron}


The \cite{Haslam1981A,Haslam1982B} all-sky map of the Galactic synchrotron emission at a wavelength of $734.8\ \mathrm{mm}$ still represents one of the most important radio surveys to this day. It is the standard which we use to validate our synthetic synchrotron emission maps generated with {\sc POLARIS}. In addition, we include $1.0\ \mathrm{mm}$ data from the Wilkinson Microwave Anisotropy Probe (WMAP) sky survey \citep[see][]{Page2007,Hinshaw2009} to acquire Stokes Q, U, and orientation maps for polarisation comparisons. These maps of intensity as well as polarisation are shown in Figure \ref{fig:AllSkyPlanck}.

To simulate all-sky maps of different observers position within the Milky Way like Au-6 galaxy we employ {\sc POLARIS} in the mode of spherical detectors using the HEALPIX pixelation scheme at a wavelength of $734.8\ \mathrm{mm}$ as well as $1.0\ \mathrm{mm}$. We do so for ten distinct observer positions as indicated in the left panel of Figure \ref{fig:MidThB}. They lie in the disc midplane at $z=0\ \mathrm{pc}$ within a radius of about $8\ \mathrm{pc} \leq r \leq 10\ \mathrm{pc}$. To best mimic the conditions in the solar neighbourhood, we select the observers to be placed within a gas density cavity similar to the Local Bubble that defines our own Galactic environment \citep[see e.g.][]{Fuchs2009,Liu2017,Alves2018}. Choosing the positions fully at random may accidentally result in an observer placed within or close to a molecular cloud. This would result in maps that are highly overshadowed by the contribution from very dense gas nearby contrary to what is observed.  

As an example of the all-sky maps generated with a {\sc POLARIS} RT calculation applying the CR1 model we show the result obtained for position P01 in Figure \ref{fig:AllSkyCR1}. The $734.8\ \mathrm{mm}$  synchrotron emission map (upper left corner) agrees well with the range of intensities found by \cite{Haslam1982B} and  presented in Figure \ref{fig:AllSkyPlanck}. However, we have stronger emission towards the Galactic Centre. This is a result of the central peak in the magnetic field component of our Galactic model (see Figure \ref{fig:ProfB}). The vectors of linear polarisation match overall the ones of the WMAP probe indicating a toroidal field component in both the Milky Way as well as our Galactic model (see also Appendix \ref{app:Toro} for a map with a purely toroidal field). Such a pattern is also known from dust polarisation observations \citep[see][]{Planck2014B}. However, both observations and synthetic maps of orientation angle $\chi$ show values closer to $-45^\circ$ above the galactic centre and $45^\circ$ below the centre whereas one would have expected $\pm 90^\circ$ along the Galactic latitude for a purely toroidal field (compare Figure \ref{fig:Toro}). We assume that this is because the signal does not probe the entire galaxy but is more dominated by the emission closer to the observer \cite[see also the discussion about the effects of the Local Bubble by][]{Alves2018}. The same holds for the maps of the Stokes Q and U component. The magnitude of synthetic Q and U emission matches with the WMAP observations, but we overpredict the  emission from the Galactic Centre.

In Figure \ref{fig:AllSkyCR2} we show maps similar to Figure \ref{fig:AllSkyCR1}, however, now the synthetic emission is based on the CR2 model. The pattern of linear polarisation is very similar to the Milky Way (Figure \ref{fig:AllSkyPlanck}) and to the CR2 model (Figure \ref{fig:AllSkyCR1}). Also the orientation map shows a similarly coherent polarisation. Comparing the lower panels of these figures reveals that the synchrotron emission in I, Q, and U, respectively, is underestimated by about one to two orders of magnitude throughout most of the galaxy. 

To explore this further and to better understand the influence of the different model parameters on the resulting synchrotron maps we quantify the spatial distribution of the emission using a multipole expansion in spherical harmonics \citep[see e.g.][for further details about this procedure]{ReisslA}. In Figure \ref{fig:SpectrumSync} we show the multipole spectrum obtained for the ten different observer positions (see Figure \ref{fig:MidThB}) in comparison to the one of the \cite{Haslam1982B} maps for the CR1 and CR2 model, respectively, with and without a turbulent component. The amplitude characterises the amount of structure in the maps on different scales, going from large to small as the multipole moments $l$ increase from low to high values. We find that all synthetic spectra exhibit the typical saw-tooth pattern well known for our Milky-Way\footnote{In fact, a saw-tooth pattern is not a fingerprint of the Milky Way but is characteristic for any disc.}. Such a pattern can also bee seen in other tracers such as $H_{\alpha}$ \citep[see e.g.][]{ReisslA}. As for synchrotron emission, the multipole spectrum of the smooth CR1 model derived from \cite{Drimmel2001} with an additional turbulent field  agrees very well with the one of the original \cite{Haslam1981A} map. There is only slight tendency for the synthetic maps to  overestimate the fluctuations in  the range  ${20 < l < 50}$ while  for $l>70$ we get a somewhat steeper slope. We note that a good match requires us to add a fluctuating magnetic field component to the relatively smooth magnetic field configuration in the underlying Au-6 galaxy. It follows a Gaussian modulation with $\sigma_B=25^{\circ}$ and $\sigma_b=2\ \rm{\mu G}$ (see Section \ref{sect:TurbulentB} for definitions). We emphasise that these particular parameters may only apply in the context of this POLARIS CR1 test setup. Different galaxies may require a different choice of parameters. They may also be degenerate with a wide range of parameters leading to similar multipole fits. Furthermore, grid artifacts may enhance the small scale structure of the synthetic maps. However, exploring this range of degeneracy goes beyond the scope of this code paper. 

When we take the magnetic field structure of Au-6 at face value and do not add a small-scale turbulent component, we still see a saw-tooth pattern for $l\leq 10$ for model CR1. However, the spectrum decays too quickly at larger $l$ indicating that the corresponding maps exhibit to little small-scale structure. We conclude that the presence of supersonic turbulence, which is ubiquitously observed in the Galactic interstellar medium and which is known to be one of the primary physical agents controlling the star formation process \citep[][]{Elmegreen2004,MacLow2004,Klessen2016}, is also important in determining the small-scale characteristics of the Galactic synchrotron emission. 

Finally, we also plot the multipole model fit for CR2 with turbulent component in Figure \ref{fig:SpectrumSync}. Yet again, we see a saw-tooth pattern pattern, but it is even less predominant and the spectrum decays even more rapidly with increasing $l$. Even with a turbulent magnetic field component, our particular CR2  model is not capable of reproducing the small-scale structures.

This may seem surprising at first sight, because CR2 exhibits much more structure than the smooth CR1 model based on \cite{Drimmel2001}. Once again, emphasises the importance of the local environment for the observed synchrotron emission. The CR2 model has large patches with little to no free electrons. If the observer is placed within such a region, as we do in P01 to P10 to mimic the Local Bubble, then the immediate surrounding will contribute very little to the observed flux. In our case, this leads to too low emission at high galactic latitutes and towards the galactic anticentre. In addition, we find too little small-scale variations resulting in a steep decline of the angular power spectrum beyond $l \approx 10$.

We emphasise that all these findings concerning the lower emission of the CR2 model may only be true for our particular Galactic disc model. In general, the CR2 model may still be an viable alternative to the CR1 parametrization presented in \cite{Drimmel2001} considering other types of MHD simulations as future inputs for {\sc POLARIS}.

\subsection{Extragalactic observations and Faraday depolarisation}
\label{sect:ObsDepolarisation}

\begin{figure*}
          \begin{minipage}[c]{1.0\linewidth}
           \begin{center}
              \includegraphics[width=0.49\textwidth]{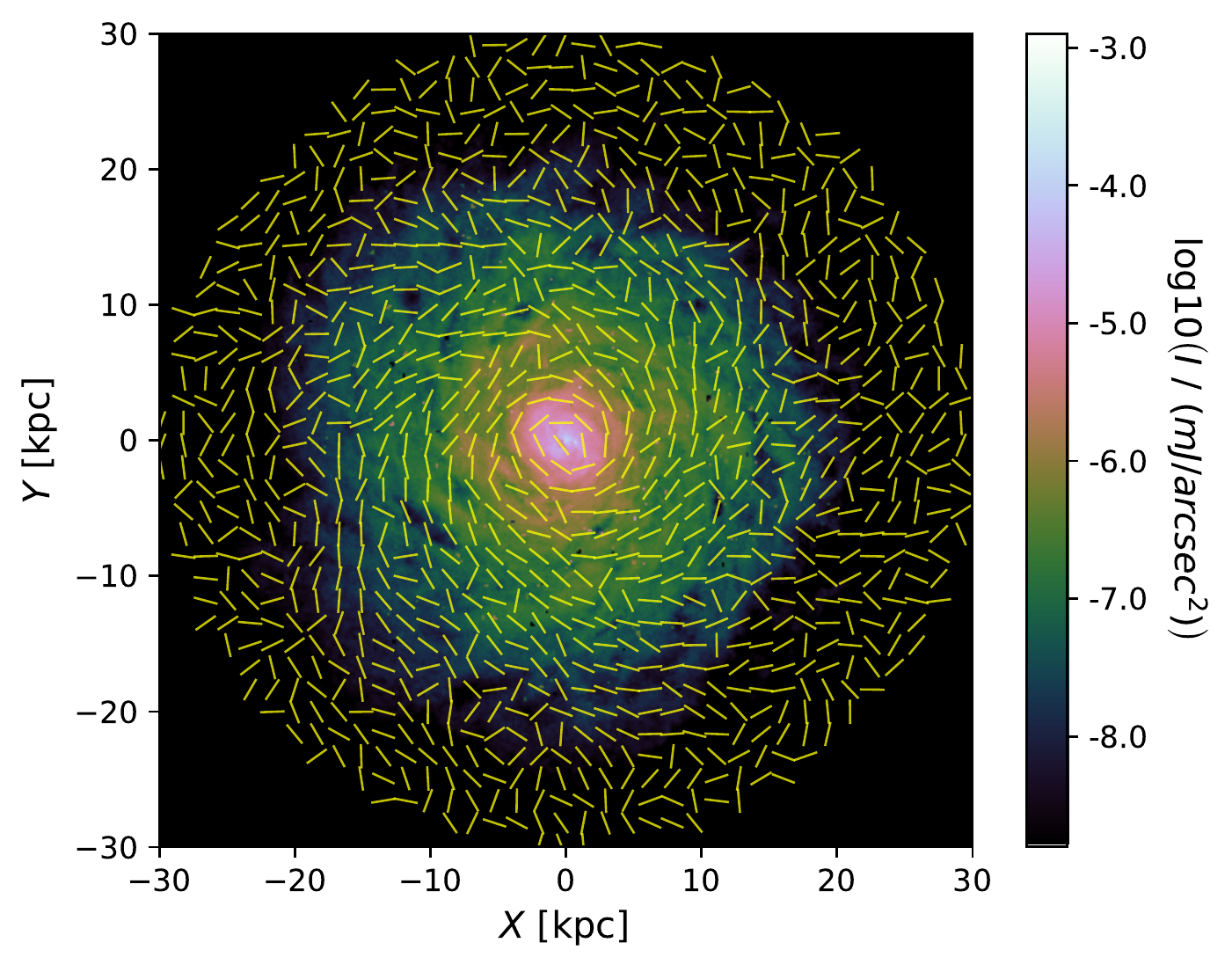}
              \includegraphics[width=0.49\textwidth]{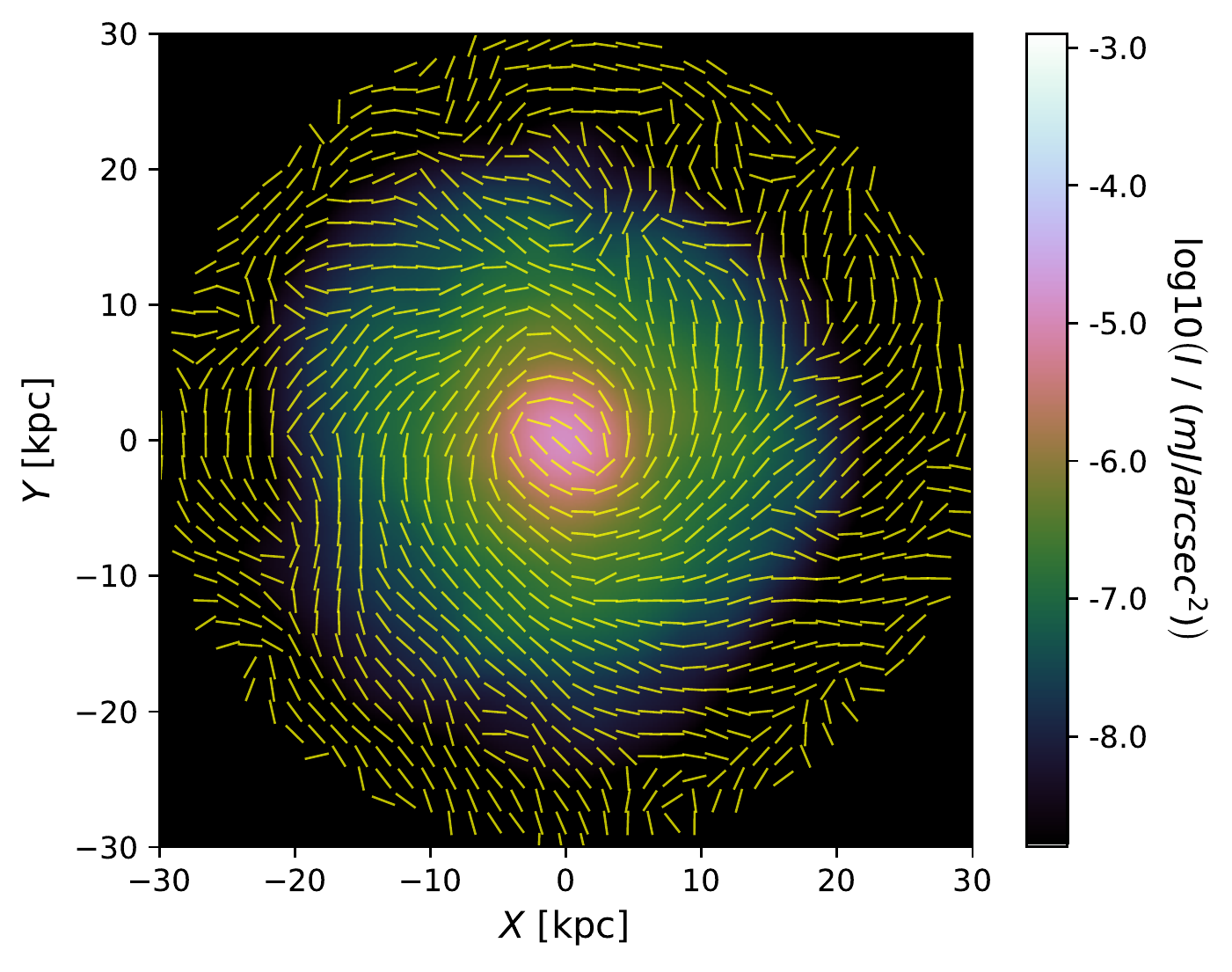}
              \includegraphics[width=0.49\textwidth]{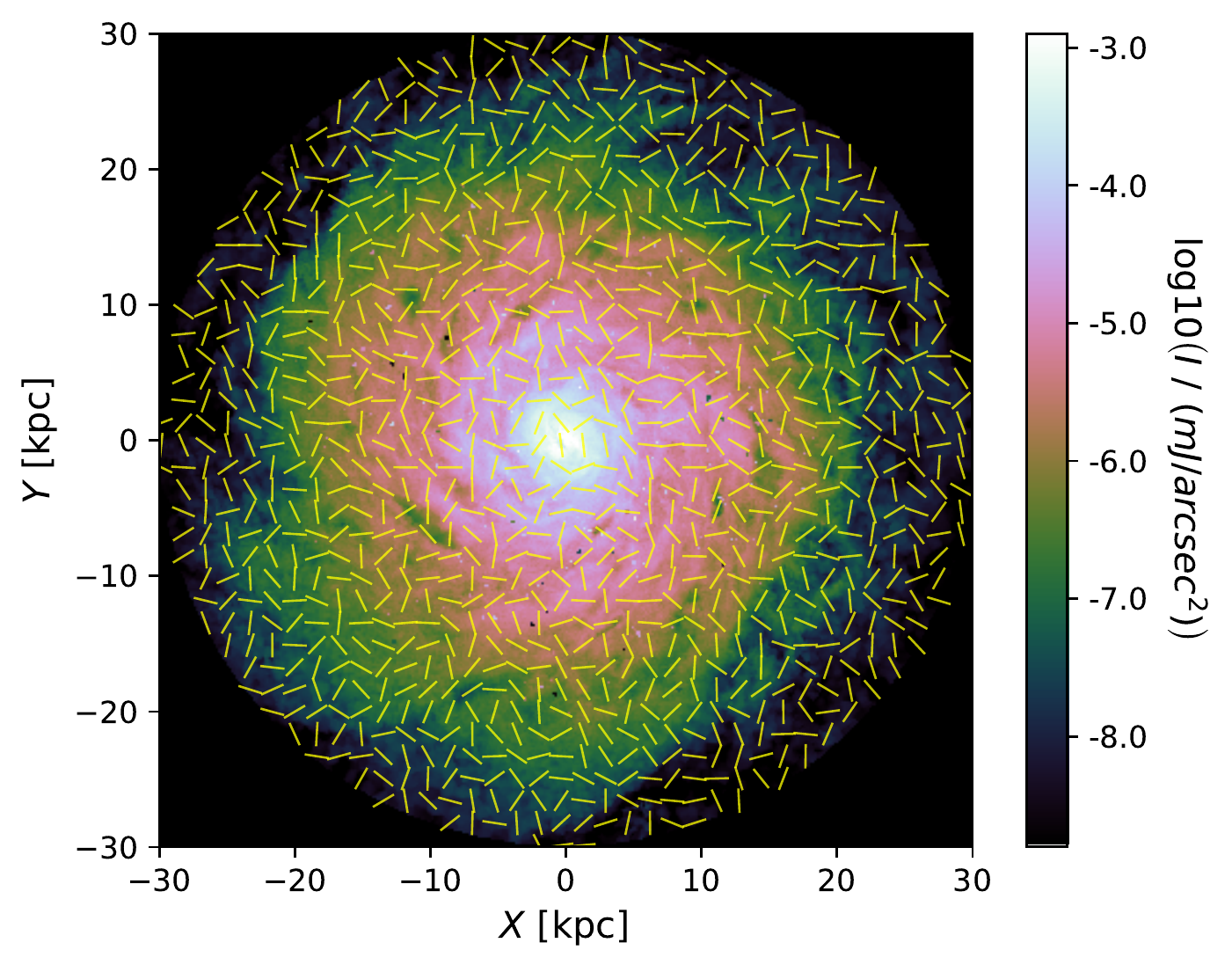}
              \includegraphics[width=0.49\textwidth]{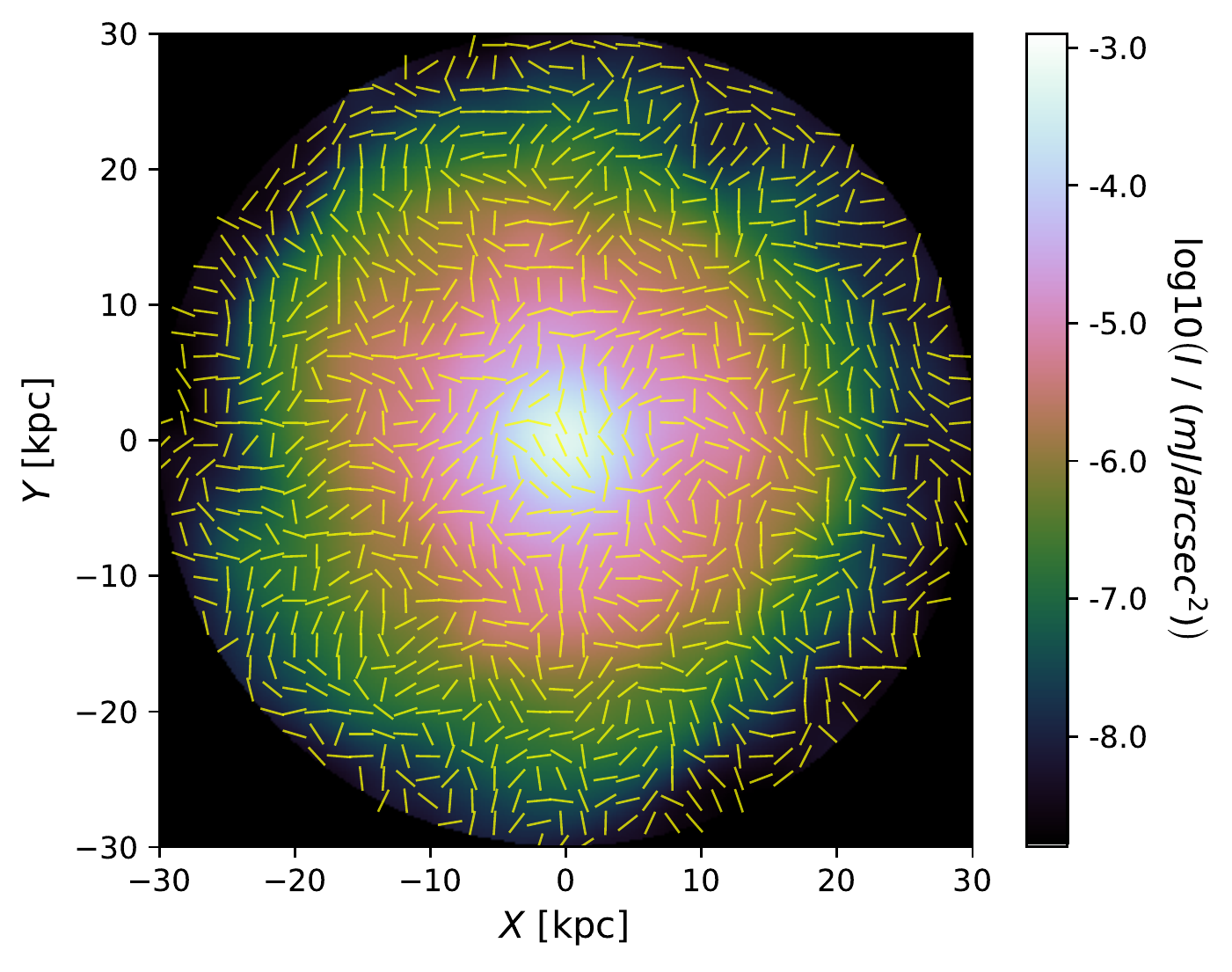}
           \end{center}
           \caption{Synchrotron emission maps for a wavelength of $62\ \mathrm{mm}$ with a resolution of $0.15''$ (top left) and $15''$ (top right) and maps for a wavelength of $201\ \mathrm{mm}$ with a resolution of $0.15''$ (bottom left) and $15''$ (bottom right). The galaxy is at a distance of $3.5\ \mathrm{Mpc}$ and overlaid with normalised polarisation vectors. The vectors are rotated by $90^\circ$ to match the actual magnetic field direction.}
	\label{fig:PlaneObsI}
        \end{minipage}
\end{figure*}
\begin{figure*}
         \begin{minipage}[c]{1.0\linewidth}
           \begin{center}
              \includegraphics[width=0.49\textwidth]{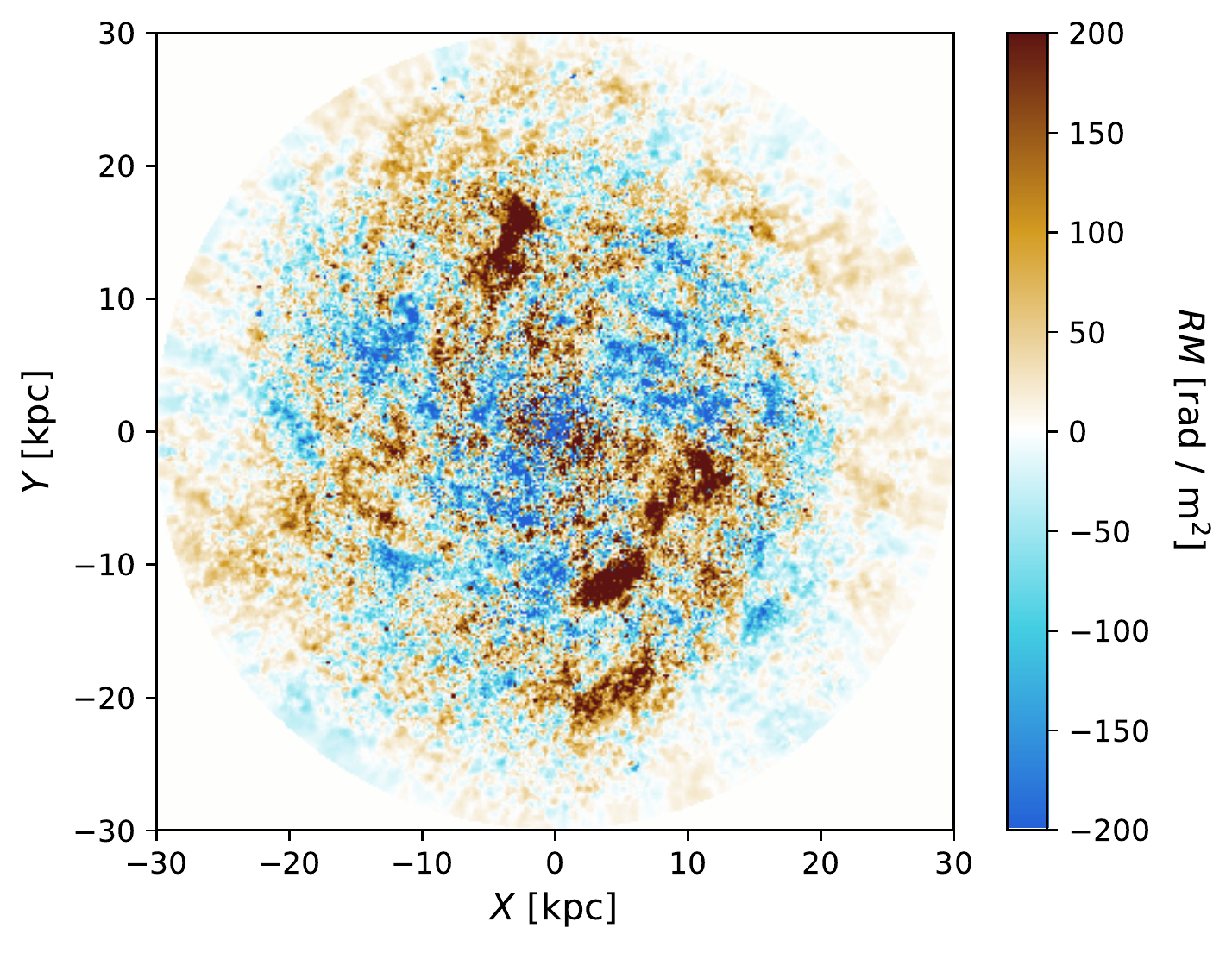}
              \includegraphics[width=0.49\textwidth]{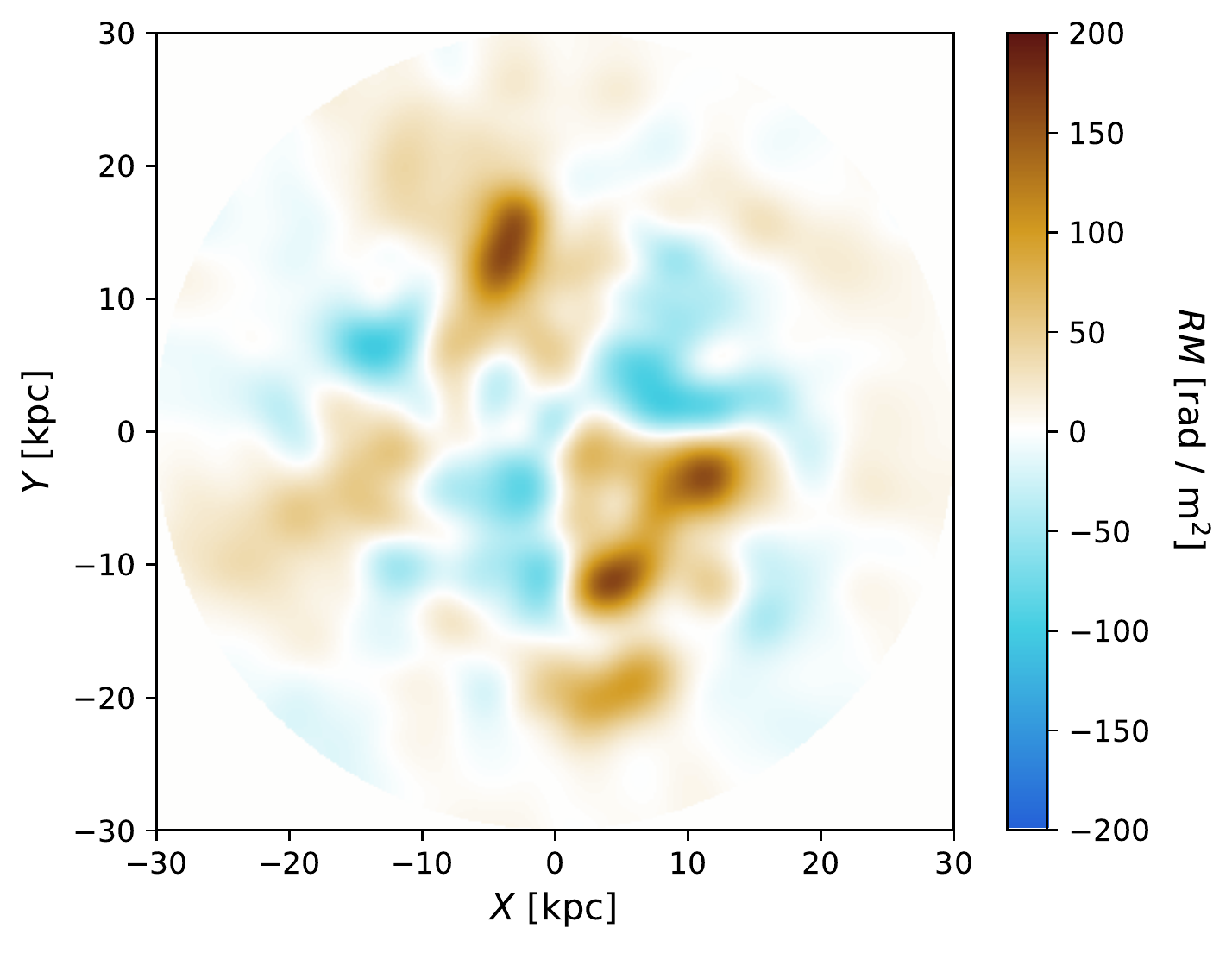}
              \includegraphics[width=0.49\textwidth]{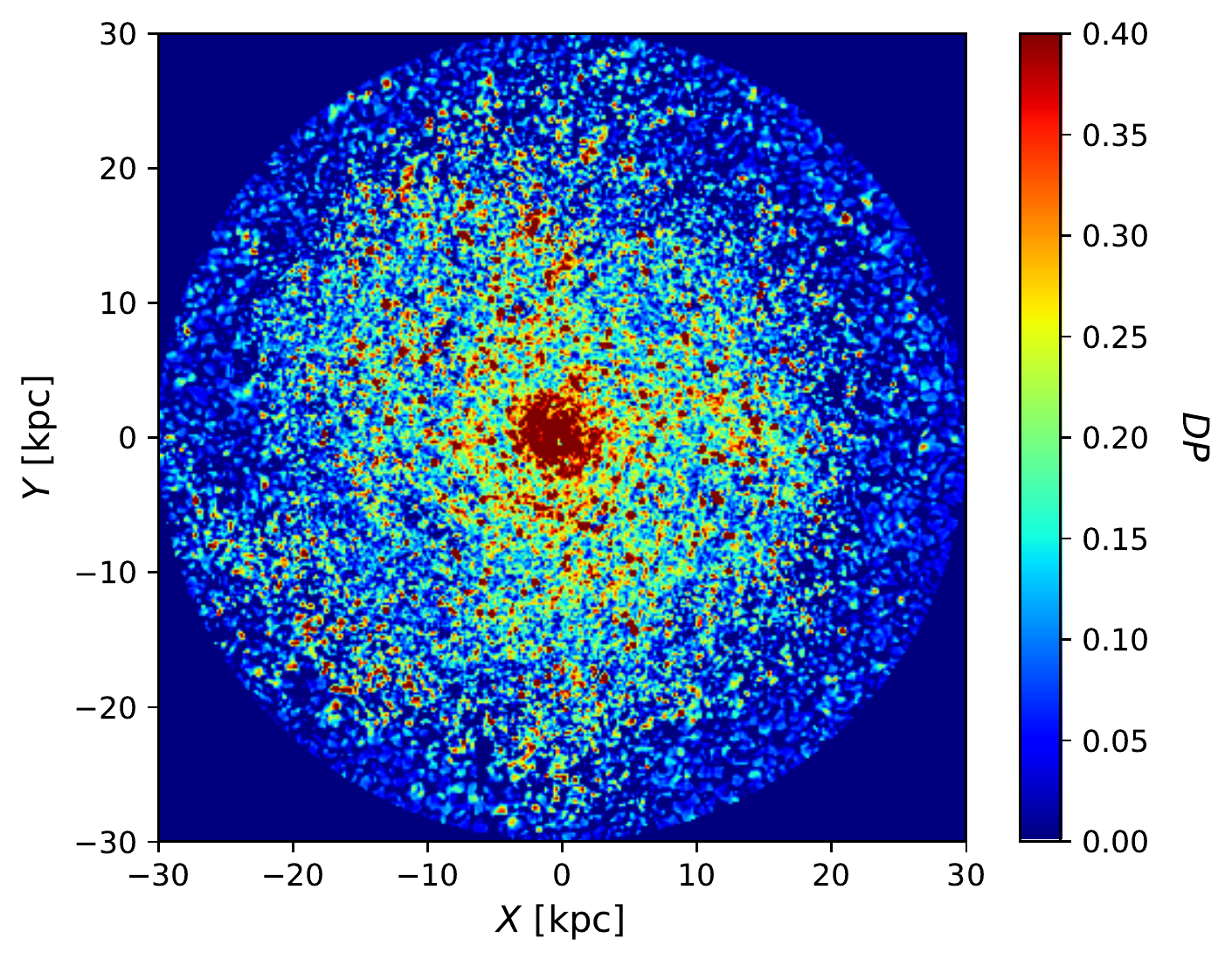}
              \includegraphics[width=0.49\textwidth]{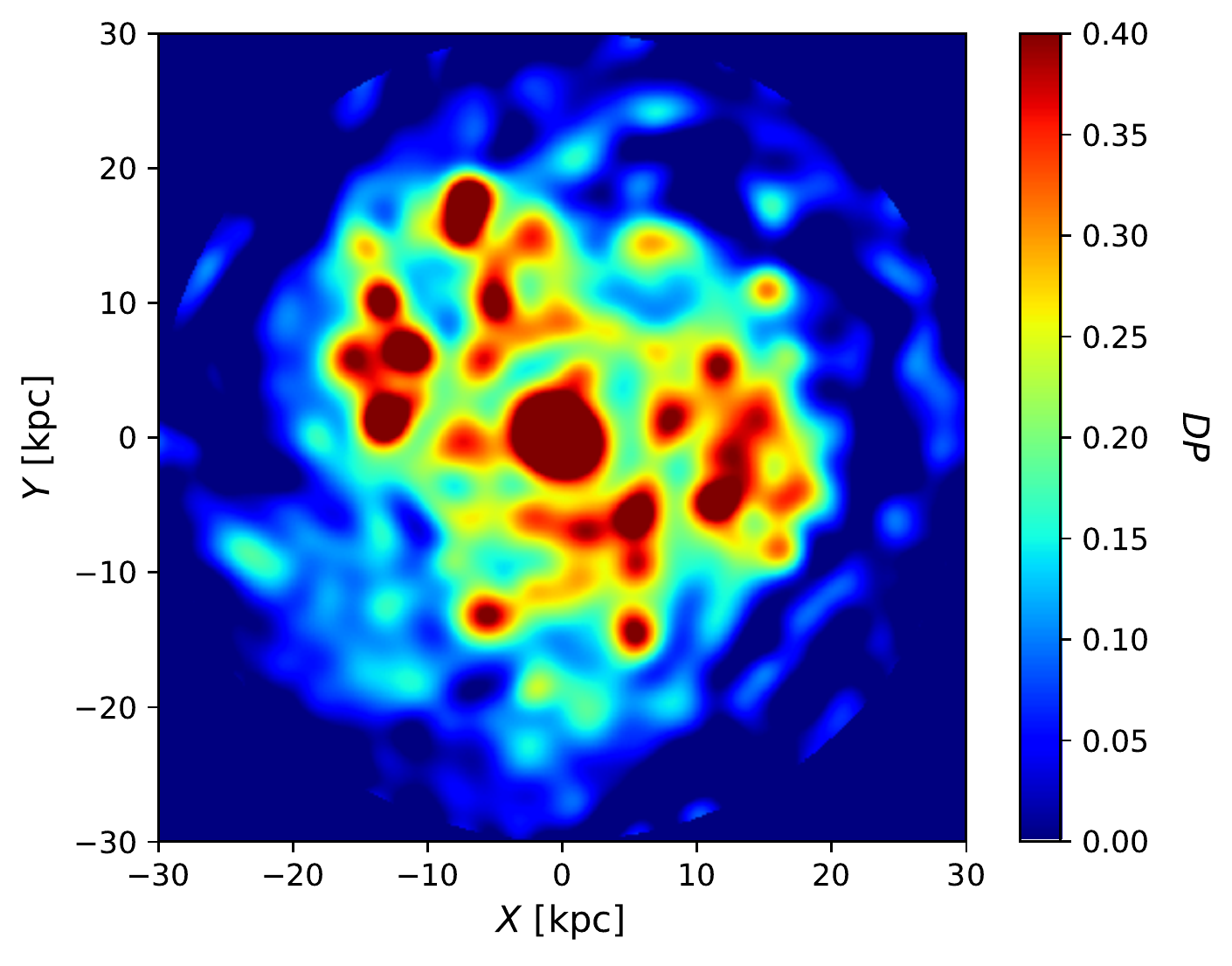}
           \end{center}
           \caption{The same as Figure \ref{fig:PlaneObsI} for the Faraday RM  (top panels) and the depolarisation fraction DP (bottom panels) as defined in Equation \ref{eq:depol} with $\lambda_1=201\ \mathrm{mm}$ and $\lambda_2=62\ \mathrm{mm}$. Again, images on the left side have a resolution of $0.15''$, while the right side was smoothed to $15''$.}
	\label{fig:PlaneObsRMDP}
        \end{minipage}

\end{figure*}
As part of the radiative transfer calculations that form the base of the emission maps discussed in the previous section {\sc POLARIS} automatically produces the corresponding map of the Faraday rotation measure (RM). 
Arguing from the Stokes vector formalism, even a parallel magnetic field may not achieve the highest expected degree of linear polarisation (see Equation \ref{eq:MaxPl}). The permanent mixing of the Q, U, and V component by means of FC and FR  may lead to a depolarisation of radiation due to  the additive nature of the Stokes vector. These depolarisation effects especially the Faraday rotation measures (RM) are extensively studied in \cite{Sokoloff1998} and observed in the nearby spiral galaxies IC 342, M51, and NGC 25, respectively, by \cite{Heesen2011,Heesen2011X}, \cite{Fletcher2011}, and \cite{Beck2015}.

In order to test the {\sc POLARIS} code for accurate predictions of depolarisation effects and RM in extragalactic objects, we produce similar observations  with a face-on planar detector at a distance of $3.5\ \mathrm{Mpc}$ away from the CR1 model with turbulent magnetic field for wavelengths of $62\ \mathrm{mm}$, $201\ \mathrm{mm}$, and $734.8\ \mathrm{mm}$. Here, resolution effects are studied by smoothing the Stokes I, Q, and U maps with a Gaussian beam of $0.15''$ and $15''$, respectively. The later resolution is comparable to that of \cite{Fletcher2011} and \cite{Beck2015}.

We remind the reader that the density of thermal and cosmic ray electrons in the outskirts of the Au-6 galaxy model, i.e. for $r > 15\ \mathrm{kpc}$, is higher than in the Milky Way (see Figure \ref{fig:ProfDens}) and so the resulting synthetic maps may not provide a fully appropriate outside view onto the Milky Way.

Figure \ref{fig:PlaneObsI} presents the resulting emission maps at $62\ \mathrm{mm}$ and $201\ \mathrm{mm}$ for different resolutions overlaid with normalised polarisation vectors tracing the magnetic field direction. For $62\ \mathrm{mm}$ and $0.15''$ the polarisation is well ordered in the disc and centre regions with a magnetic field following the spiral pattern, but the polarisation becomes increasingly disordered towards the outer edge of the disc. For the $15''$ smoothed map, the toroidal component becomes even more apparent. Such a pattern seems to be a common feature in radio and optical spiral disc observations \citep[e.g.][]{Fendt1998,Soida2002,Fletcher2011,Beck2015,Frick2016} indicating again a strong toroidal field with a non-negligible turbulent component. Observing at $201\ \mathrm{mm}$ (or even $734.8\ \mathrm{mm}$) leads to severe perturbations of this coherent pattern even in the centre, since FR becomes increasingly dominant (see also Equation \ref{eq:ChiRot} and Figure \ref{fig:PlaneObsRMDP}). Here, synchrotron polarisation does no longer allow to accurately trace the magnetic field morphology at longer wavelengths. A similar trend was observed in the polarisation pattern of M51 for $62\ \mathrm{mm}$ and $201\ \mathrm{mm}$ as presented in \cite{Fletcher2011}. In contrast to the $201\ \mathrm{mm}$ maps in Fig. \ref{fig:PlaneObsI}, in M51  observations the correlation between the spiral arms and the polarisation patters is not completely lost. However, M51 has only two spiral arms with large inter-arm regions while the Au-6 is much more tightly wound. Hence, the $201\ \mathrm{mm}$ polarisation pattern in Fig. \ref{fig:PlaneObsI} can no longer clearly be attributed to any particular spiral even in our high resolution map.

The corresponding map of the Faraday RM is depicted in the top panels of Figure \ref{fig:PlaneObsRMDP}. The spatial distribution as well as the magnitude of the effect (up to $\pm 200\ \mathrm{rad/m^2}$) match well with the expectations from a similar high-resolution RM study based on the original AU-6 MHD \citep{Pakmor2018} and with what is known from extragalactic observations \citep[e.g.][]{Heesen2011,Fletcher2011,Beck2015}. Faraday rotation can be the result of density fluctuations within the thermal electron distribution as well as in the magnitude and orientation of the magnetic field. Because the magnetic field structure of the Au-6 galaxy is rather regular, even when adding a turbulent component, we conclude that the changes in the magnitude of RM seen in the top panels of Figure \ref{fig:PlaneObsRMDP} is mostly due to fluctuations in the thermal electron density. However, this clearly deserves further investigation. Any follow-up studies would also need to take the contribution of the halo field to the total RM into account, which is an effect that we neglect here. 

In the bottom panels of Figure \ref{fig:PlaneObsRMDP} we also show a map of the depolarisation fraction DP, as defined in Equation \ref{eq:depol}, using $\lambda_1=201\ \mathrm{mm}$ and $\lambda_2=62\ \mathrm{mm}$ as well as a constant spectral index of $\alpha = 1$. The magnitude varies mostly between $0$ and $0.4$ with peak values up to $0.8$ for both the $0.15''$ and the $15''$ maps. This result concurs with the maps presented in \cite{Beck2015}, \cite{Fletcher2011} and \cite{Heesen2011X}, although the latter authors also report peak values up to unity. Our map does not particularly resemble the spiral structure of the emission with most of the DP occurring in distinct spots. We note that, these spots are connected to the most ionizing cluster regions of the population synthesis model of \cite{ReisslA}. The lack of correlation between density structures and DP is also consistent with observations e.g. with the M51 DP map presented in \cite{Fletcher2011}.

We emphasize that the native resolution of our synthetic extragalactic observations corresponds to $0.15\ \mathrm{''}$, which is a hundred times better than the data presented in \cite{Heesen2011,Heesen2011X}, \cite{Fletcher2011}, or \cite{Beck2015}, respectively. This demonstrates the high quality of the data coming from the {\sc POLARIS} RT simulations.

\section{Summary}
\label{sect:Summary}

In this paper, we presented an updated and extended version of the polarisation radiative transfer code {\sc POLARIS}. The new code solves the full four Stokes parameters matrix equation of the radiative transfer problem in order to create synthetic synchrotron emission maps including polarisation, Faraday conversion and Faraday rotation. {\sc POLARIS} can be used to generate synthetic observations based on multi-physics numerical simulations as input running on the native grids of all major astrophysical MHD codes. 

As a case study, we tested the accuracy and predictive capability of the {\sc POLARIS} code through a set of radiative transfer simulations based on the  Auriga cosmological MHD zoom simulation project \cite[][]{Grand2017}. The selected galaxy, Au-6, is an analog of the Milky Way. We modified it by employing the star cluster population synthesis model WARPFIELD-POP presented by \citep[][]{ReisslA} to produce a more realistic distribution of thermal electrons and by adding a turbulent component to the original magnetic field. To explore the impact of cosmic ray electrons, we investigated to different approaches based on exiting models. The radiative transfer simulations we ran explored the influence of the different post-processing steps, electron distributions, as well as observational conditions of the Auriga galaxy on synchrotron observables. We focused our attention on those wavelengths that are most commonly used in observations of Galactic magnetic fields ($1\ \mathrm{mm}$ to $730\ \mathrm{mm}$). Our synthetic synchrotron all-sky maps match well with actual observations both in magnitude and structure. Furthermore, we produced and examined mock observations of extragalactic systems, which show familiar patterns in polarisation, Faraday rotation measure, and depolarisation. Altogether, we demonstrated that {\sc POLARIS} is a tool that reliable computes synchrotron emission, polarisation, the internal and external depolarisation and Faraday rotation effects. It can produce reliable all-sky maps for a fictitious observer within a galaxy and it can create images of galaxies as seen from far away. {\sc POLARIS} is a highly versatile radiative transfer code that can be used for the detailed comparison with Galactic and extragalactic observations. 

We summarise our scientific findings as follows: (i) Different methods to derive Galactic cosmic ray electron distributions, based on a simple parametrisation and on energy equipartition, reproduce the observed synchrotron polarisation pattern. However, the equipartition approach seems to underestimate the total amount of synchrotron emission. (ii) The presence of a turbulent magnetic field component is required to reproduce the observed Galactic small-scale structures of the synchrotron emission. (iii) Our radiative transfer simulations indicate that the observed Galactic synchrotron emission depends strongly on the actual position within the Milky Way disc. (iv) The depolarisation pattern observed in our synthetic galaxy by an observer far away is largely accounted for by Faraday rotation, its small-scale features are mostly dominated by the thermal electron distribution.

In a series of forthcoming papers we plan to utilise {\sc POLARIS} to further minimise observed ambiguities in magnetic field measurements by means of dust polarisation \cite[e.g.][]{Reissl2014,Reissl2018} or Zeeman effect \cite[][]{Brauer2017,Brauer2017B,Reissl2018}. The set of unique polarisation features unified in a single code has also the potential to address open questions concerning the separation the CMB measurements from the pollution of dust and synchrotron polarisation coming from our own Milky-Way.

\appendix
\section{Error estimation and code limitations}
\label{app:Error}

\begin{figure*}
 \begin{center}
         \begin{minipage}[c]{1.0\linewidth}
           \begin{center}
              \includegraphics[width=0.49\textwidth]{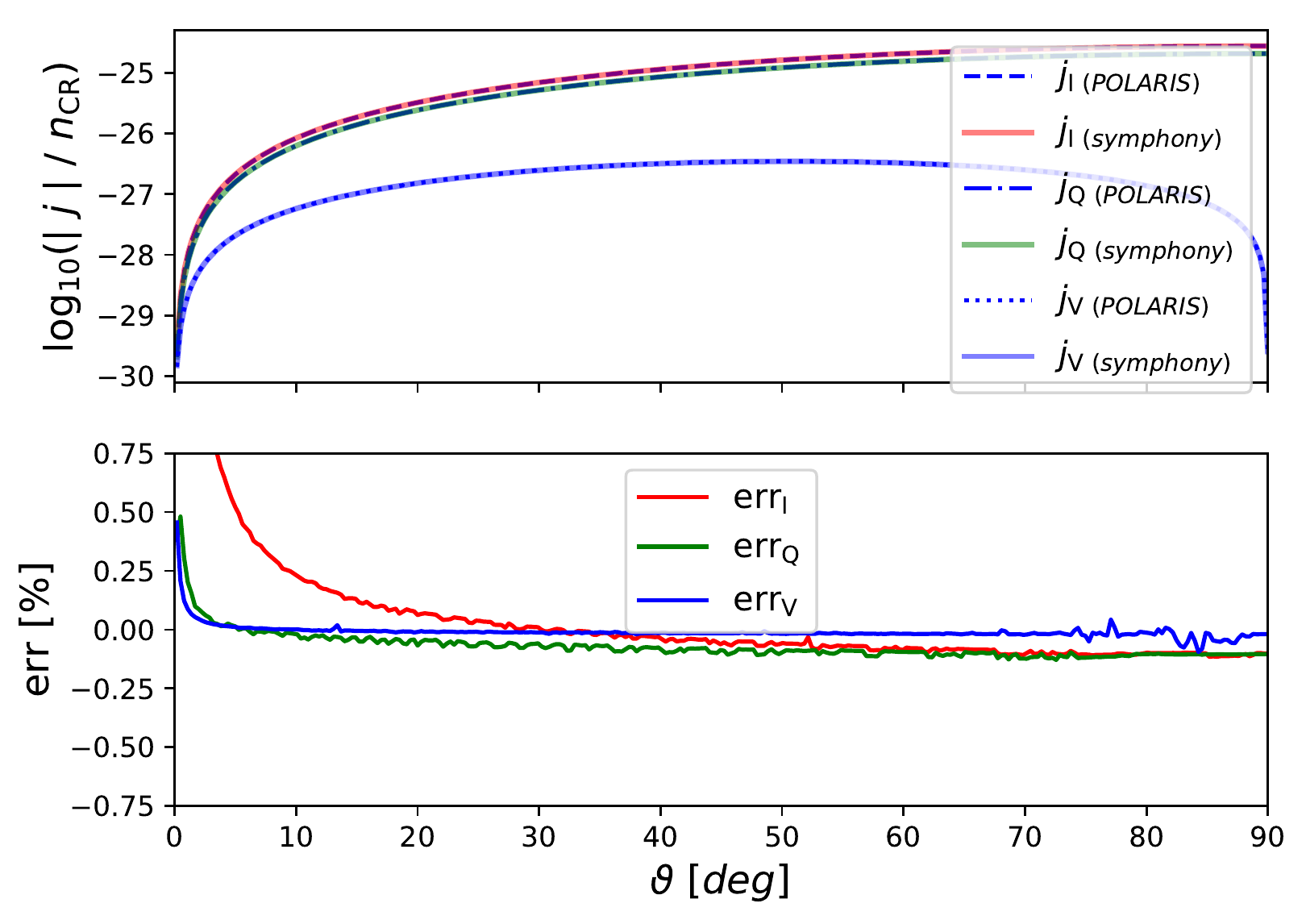}
              \includegraphics[width=0.49\textwidth]{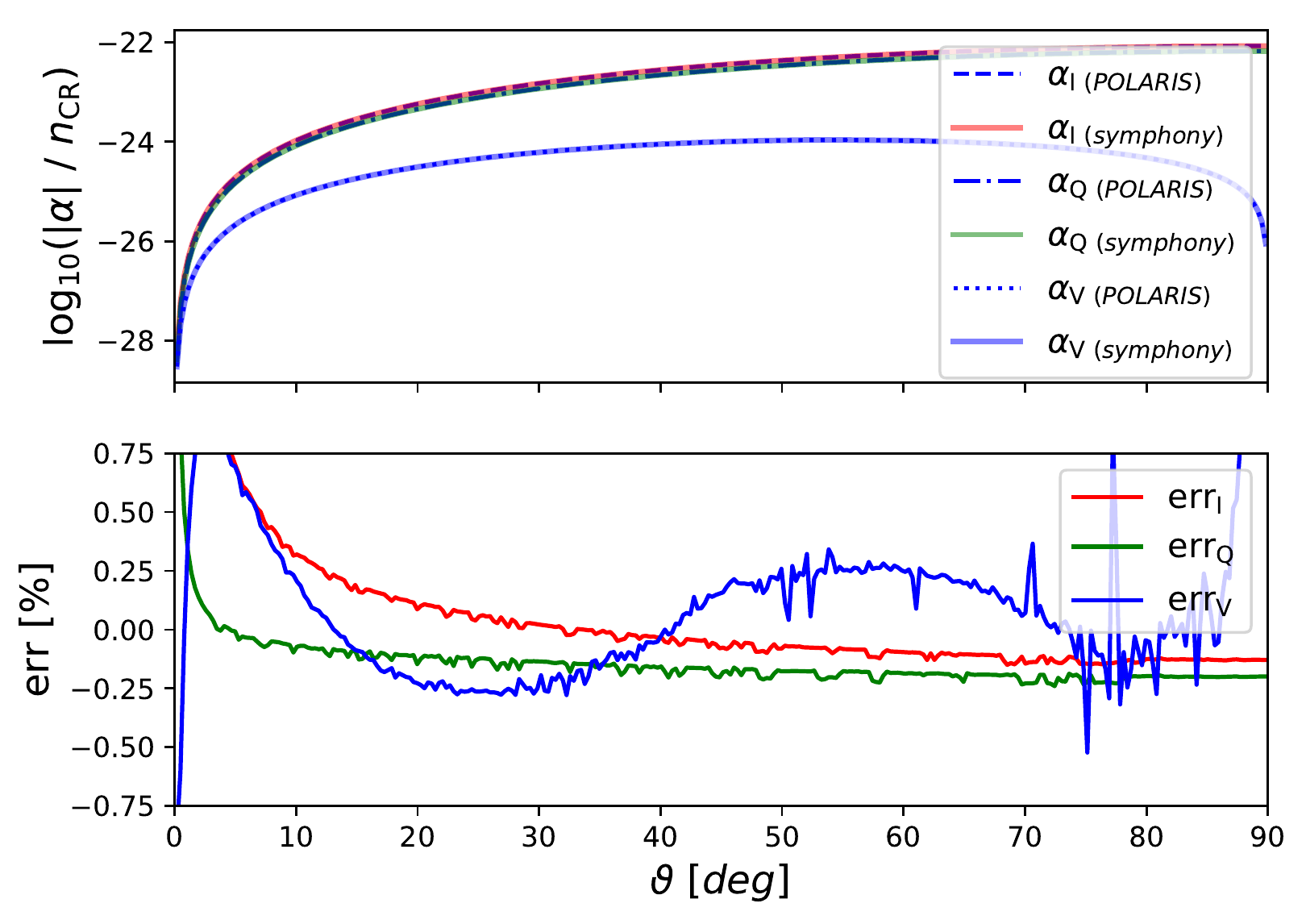}
           \end{center}
           \caption{Left column: Fitted emission coefficient  $j_\mathrm{I}$ for intensity,  $j_\mathrm{Q}$ for linear polarisation, and  $j_\mathrm{V}$ for circular polarisation as they are implemented in {POLARIS} over angle $\vartheta$ in comparison with exact integral solutions provided by the {SYMPHONY} code. Here, the parameters are $\lambda=734.8\ \mathrm{mm}$, $B=30\ \mathrm{G}$, power-law index $p=3$,  $\gamma_{\mathrm{min}}=4$, and $\gamma_{\mathrm{min}}=300$. The error is defined to be $\mathrm{err}=1-j_{(\mathrm{{POLARIS}})}/j_{(\mathrm{{SYMPHONY}})}$. Right column: The same as the left column for the corresponding emission coefficients $\alpha_\mathrm{I}$, $\alpha_\mathrm{Q}$, and $\alpha_\mathrm{U}$, respectively. Note, that all Q coefficients are negative and all V coefficients do change their sign from positive to negative values for $\vartheta>90^{\circ}$.}
	\label{fig:Errors30Gauss}
        \end{minipage} 
        
        \begin{minipage}[c]{1.0\linewidth}
           \begin{center}
              \includegraphics[width=0.49\textwidth]{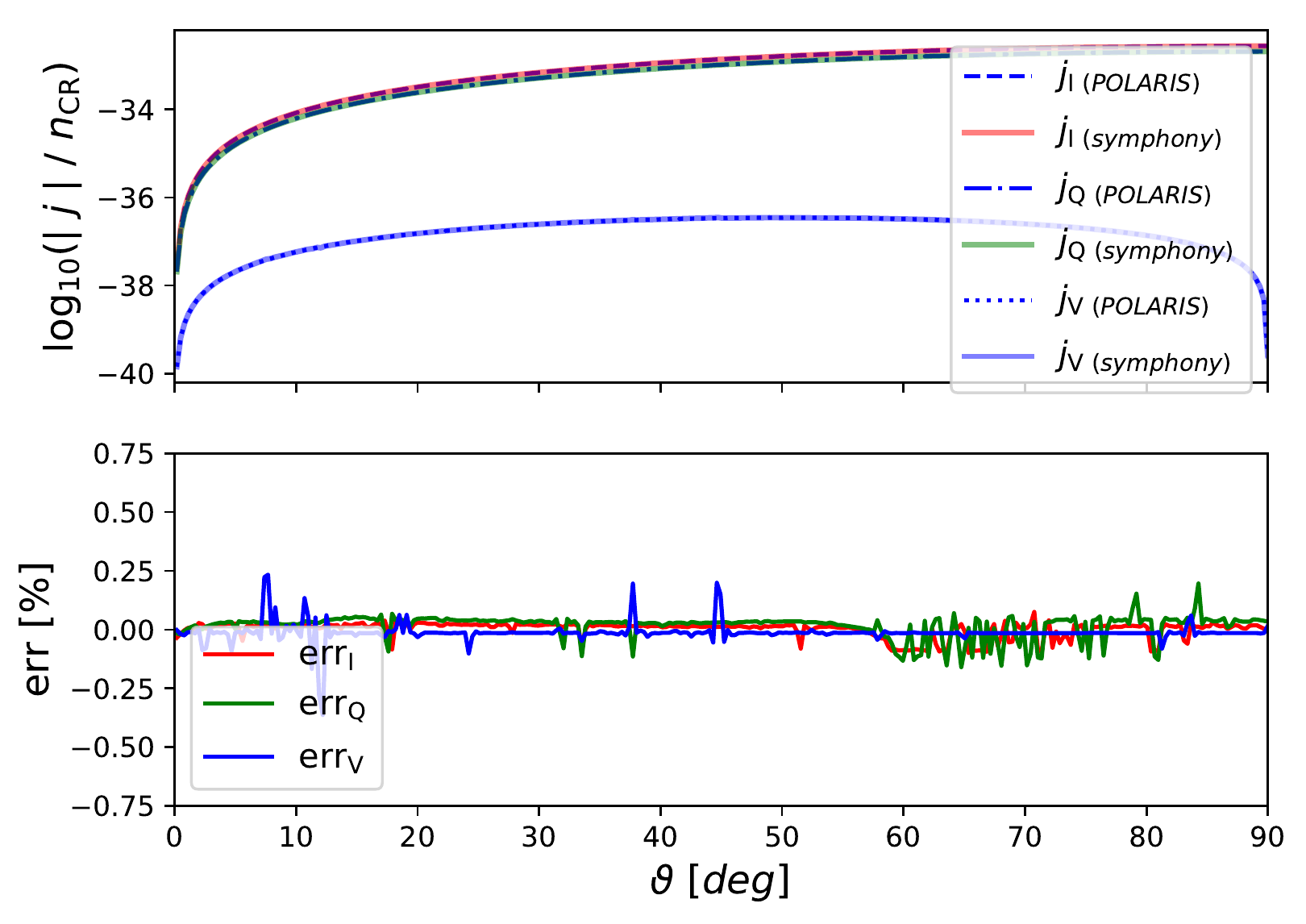}
              \includegraphics[width=0.49\textwidth]{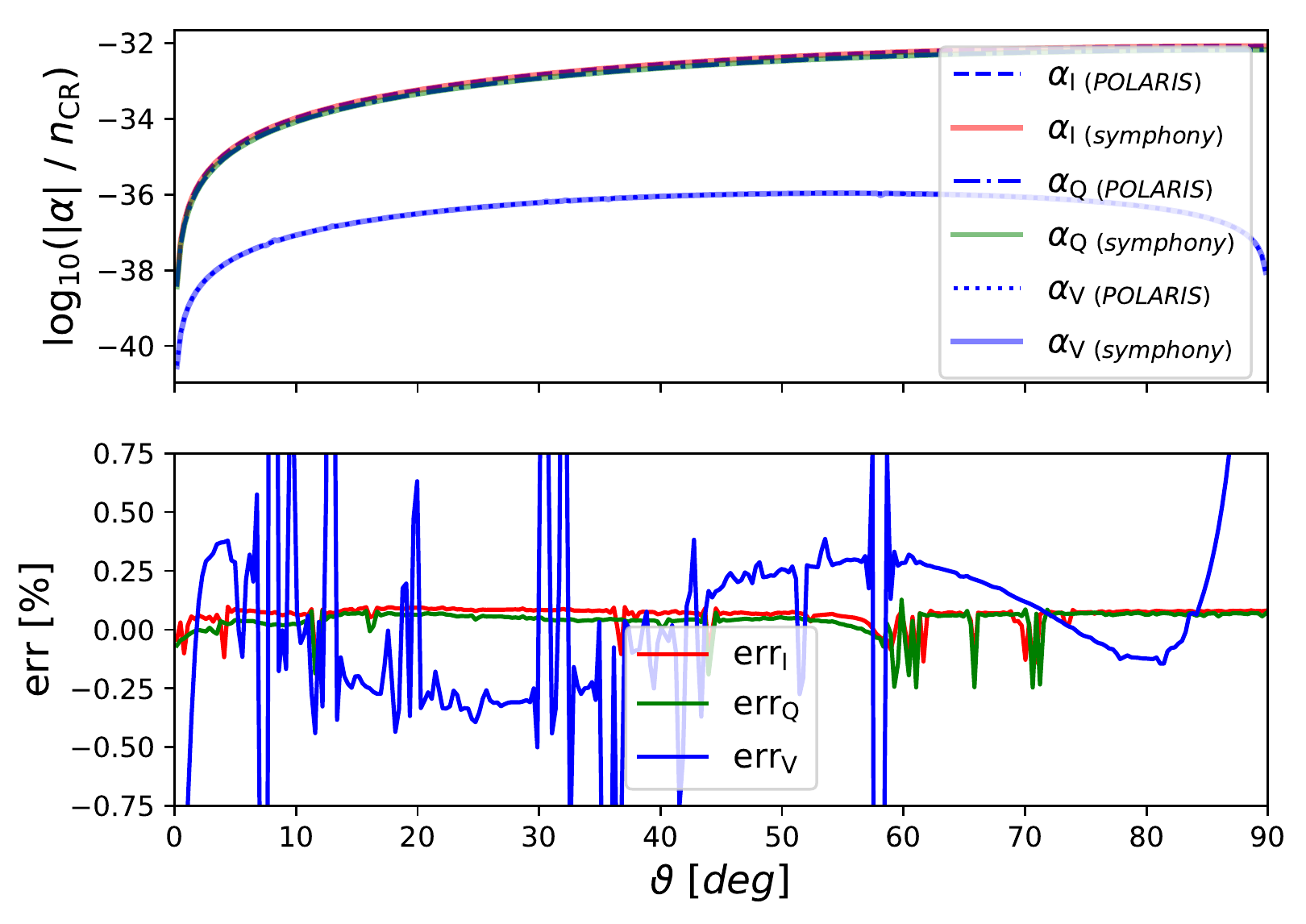}
           \end{center}
           \caption{The same as Figure \ref{fig:Errors30Gauss} for $B=3\ \mathrm{mG}$.}
	\label{fig:Errors0.003Gauss}
        \end{minipage} 

\end{center}  

\end{figure*}
In this section we explore the limitations of the applied fit functions in comparison with the exact integral solutions of the coefficients of synchrotron emission, absorption, Faraday conversion (FC), and Faraday rotation (FR). For that we consider an maximum error of $1\ \%$ to be acceptable for the {\sc POLARIS} implementation. This limits the range of wavelength $\lambda$ available for synthetic observations to ${10^2 < \lambda_{\rm c}/\lambda < 10^9}$ (see Equation \ref{eq:LambdaC} for the definition of $\lambda_{\rm c}$). In turn the range of the energy spectrum is given by ${\gamma_{\rm min}^2\ll\lambda_{\rm c}/\lambda < \gamma_{\rm max}^2}$ \citep[see][for details]{Pandya2016}. We note that the upper limit is less strict as long as the ratio ${\lambda_{\rm c}/\lambda}$ is of the same order as $\gamma_{\rm max}^2$. 

Errors up to $35\ \%$ are reported in \cite{Pandya2016}. However, these uncertainties are given for a magnitude of the magnetic field in the order of $\approx 10\ \mathrm{G}$ and for small values of $\vartheta$. Indeed, this rather high field strength is far beyond values typical for the Milky Way where we can expect values of about $6\ \mathrm{\mu G}$ \citep[][]{Strong2000} in our local environment and up to $1\ \mathrm{m G}$ \citep[][]{YusefZadeh1996} near the Galactic Centre. A similar range of field strengths can be expected for other spiral galaxies \citep[][]{Niklas1995}.

We compare the integral solutions calculated with the {\sc SYMPHONY} code \citep[][]{Pandya2016} to the fit functions as they are implemented in {\sc POLARIS}. Here, we find for $B\ll 1\ \mathrm{G}$ the absorption and emission coefficients are less prone to errors even for smaller values of the angle $\vartheta$ (compare Figures \ref{fig:Errors30Gauss} and \ref{fig:Errors0.003Gauss}). Consequently, we can apply the fit functions throughout the parameter space of the Milky Way model presented in this paper. 

We report some irregularities between fit functions and exact integral solutions for a lower cut-off of $\gamma_{\mathrm{min}} \neq 1$. Indeed, the fit functions in {\sc SYMPHONY} do not apply for $\gamma_{\mathrm{min}}>1$\footnote{This fact was confirmed by Alexander Pandya via private conversation.}. Hence, we corrected Equations \ref{eq:CRJI} and \ref{eq:CRAlphaI} by an additional factor of $\gamma_{\mathrm{min}}^{1-p}$ in order to provide general solutions for any $\gamma_{\mathrm{min}}>1$.

We also find an offset between the integral solution and the fit functions of the absorption coefficient $\alpha_{\mathrm{Q}}$ of about $4\ \%$ being constant over a large parameter range. Assuming the integral solutions to be correct we multiply an additional factor of $996/1000$ to decrease this offset in Equation \ref{eq:CRAlphaQ}.

A similar problem occurs for the absorption coefficient $\alpha_{\mathrm{V}}$. Here, the offset depends on the angle $\vartheta$ and exceeds the demanded $1\ \%$ error limit over a wide range of $\vartheta$. We apply an additional correction function to Equation \ref{eq:CRAlphaV} defined to be  
\begin{equation}
      k_{\mathrm{V}}(\vartheta) = \begin{cases} 0.9914 + 0.0075\times\vartheta^{11/12}    &\mbox{if } \vartheta \leq 0.8034 \\0.9919 +  0.0013/\sin(0.0048 + \vartheta)  & \mbox{else}  \end{cases}\, .
        \label{eq:QGamma}
\end{equation}
In Figures \ref{fig:Errors30Gauss} and \ref{fig:Errors0.003Gauss} we plot the emission and absorption as well as the errors between the implementation in {\sc SYMPHONY} and {\sc POLARIS}. We attribute the noise in the error plots to the integration scheme of {\sc SYMPHONY}. Indeed, for $B \leq 3\ \mathrm{mG}$ the implemented fit functions in {\sc POLARIS} agree very well with the integral solutions of {\sc SYMPHONY}. The only exception is the corrected absorption coefficient of circular polarisation $\alpha_{\mathrm{V}}$. The correction function pushes the error bellow $0.75\ \%$ except for angles $\vartheta<3^\circ$ and $\vartheta>87^\circ$, respectively. However, we consider this error acceptable since $\alpha_{\mathrm{V}}\rightarrow 0$ much faster as $\alpha_{\mathrm{I}}$ and $\alpha_{\mathrm{Q}}$, respectively, in this range and contributes only marginally to the total RT process.

Furthermore, the implemented FR and FC coefficients are only valid in a regime with electron temperatures of $T_{\mathrm e} \ll 10^{9}-10^{10}\ \mathrm{K}$. In a more extreme environment such as black hole accretion flows \citep[see e.g.][]{Rajesh2010,Chael2017} the  coefficients in Equation \ref{eq:kappaV} and Equation \ref{eq:kappaQ} need to be modified by some additional correction factors as discussed e.g. in \cite{Shcherbakov2008} or \cite{Dexter2016}.

\section{Numerical solver and adaptive step size}
\label{app:Solver}
Implemented in the {\sc POLARIS} code is a Runge-Kutta-Fehlberg (RFK45) solver in order to provide a high accuracy solution to the matrix RT problem. This method uses an inbuilt step size $d\ell$ correction to keep the error below a certain threshold $\epsilon$. Here, the solver compares the fourth order Runge-Kutta solution $X_{I,Q,U,V,4}$ for any of the Stokes component with the fifth order solution $X_{I,Q,U,V,5}$ by
\begin{equation}
	\epsilon_\mathrm{I,Q,U,V} = \left| \frac{X_{I,Q,U,V,4} - X_{I,Q,U,V,5}}{\epsilon_\mathrm{err}X_{I,Q,U,V,5}+\epsilon_\mathrm{abs}}   \right|\, .
\end{equation}
By default the relative error is implemented in {\sc POLARIS} to be ${\epsilon_\mathrm{err}=10^{-8}}$ with an absolute error of ${\epsilon_\mathrm{abs}=10^{-30}}$ . The RFK45 method solves the RT problem usually within a few steps per cell. However, in RT with synchrotron polarisation we have to handle the permanent transfer between the Q, U, and V components via the FR and FC coefficients. Consequently, the system of differential equations oscillates between these Stokes parameters of polarisation. A step size that is only based on the I parameter might be too large for the other Stokes parameter and can lead to the forbidden condition of $I<p_\mathrm{t}$. Hence, we account for this case by calculating four separate thresholds $\epsilon_\mathrm{I,Q,U,V}$ for each of the Stokes parameters.
The final threshold is then
\begin{equation}
	\epsilon = \min\left(\epsilon_{\mathrm{I}}, \epsilon_{\mathrm{Q}}, \epsilon_{\mathrm{U}}, \epsilon_{\mathrm{V}}\right)\, .
\end{equation}
For the case of $\epsilon>1$ a smaller step size $d\ell_\mathrm{new}$ for the current integration step needs to be determined according to
\begin{equation}
	d\ell_\mathrm{new} = \min\left(\frac{1}{10} \times d\ell_\mathrm{old},\frac{1}{4}  d\ell_\mathrm{old} \epsilon^{-0.2} \right)\, .
\end{equation}						
Otherwise, for $\epsilon\leq 1$ the integration step $d\ell$ is sufficiently small to solve the equation system of synchrotron RT within the defined error limits. Finally, the simulations stops when a all rays have reached the detector. 

We note that, under rare conditions the solver may still need several thousand steps within a singles cell. In order to circumvent this problem we implemented an alternative solver scheme. When the number of steps per cell exceeds a number of $5\times10^{5}$ we separate the RT problem. Since the FR and FC coefficients usually require the smaller $d\ell$ we write the system of equations as 
\begin{equation}
\frac{d}{d\ell}\vec{S}=-\left( \hat{K}_\alpha+\hat{K}_\kappa  \right)\vec{S}+\vec{J}\, ,
\label{eq:RTProblem3}
\end{equation}
where
\begin{equation}
\hat{K}_\alpha=\begin{pmatrix} 
\alpha_{\rm I} & \alpha_{\rm Q} & 0 & \alpha_{\rm V} \\  
\alpha_{\rm Q} & \alpha_{\rm I} & 0 & 0 \\ 
0 & 0 & \alpha_{\rm I} & 0 \\ 
\alpha_{\rm V} & 0 & 0 & \alpha_{\rm I} \end{pmatrix}
\label{eq:RTProblem3}
\end{equation}
is the absorption matrix and
\begin{equation}
\hat{K}_\kappa=\begin{pmatrix} 
0 & 0 & 0 & 0 \\  
0 & 0 & \kappa_{\rm V} & 0 \\ 
0 & -\kappa_{\rm V} & 0 & \kappa_{\rm Q} \\ 
0 & 0 & -\kappa_{\rm Q} & 0 \end{pmatrix}
\label{eq:RTProblem4}
\end{equation}
is the Faraday matrix. Now, we solve only the RT problem with $\hat{K}_\alpha$ by means of the RFK45 solver and $\epsilon_\mathrm{I}$ leading to a step size $d\ell_\mathrm{I}$ and finally to a solution of $\vec{S}_{\alpha}$ ignoring FR and FC effects.
For the Faraday part of the equation we make use of the analytically solution derived by \cite{Dexter2016}. Considering only $\hat{K}_\kappa$ the oscillation of the Stokes polarisation parameters Q, U, and V by means of Faraday mixing can analytically be calculated as
\begin{equation}
Q_\kappa=\frac{\kappa_Q}{\kappa^2} \left( j_Q\kappa_Q + j_V\kappa_V \right)  d\ell_\mathrm{I}  +\frac{\kappa_V}{\kappa^3} \left( j_V\kappa_Q + j_Q\kappa_V \right)\sin\left(\kappa   d\ell_\mathrm{I}   \right) - \frac{j_U\kappa_V}{\kappa^2}\left[ 1- \cos\left(\kappa   d\ell_\mathrm{I}   \right) \right]\, ,
\label{eq:QKappa}
\end{equation}

\begin{equation}
U_\kappa=\frac{j_Q\kappa_V-j_V\kappa_Q}{\kappa^2}\left[ 1-\cos\left(\kappa   d\ell_\mathrm{I}    \right) \right]+ \frac{j_U}{\kappa}\sin\left(\kappa   d\ell_\mathrm{I}    \right)\, ,
\label{eq:UKappa}
\end{equation}
and
\begin{equation}
V_\kappa=\frac{\kappa_V}{\kappa^2} \left( j_Q\kappa_Q + j_V\kappa_V \right)  d\ell_\mathrm{I}  +\frac{\kappa_V}{\kappa^3} \left( j_Q\kappa_V + j_V\kappa_Q \right)\sin\left(\kappa   d\ell_\mathrm{I}   \right) + \frac{j_U\kappa_Q}{\kappa^2}\left[ 1- \cos\left(\kappa   d\ell_\mathrm{I}   \right) \right]
\label{eq:VKappa}
\end{equation}
with $\kappa^2=\kappa_Q^2+\kappa_U^2$. This second set of equations results in a solution of $\vec{S}_\kappa=(0,Q_\kappa,U_\kappa, V_\kappa)^T$. The final solution is then simply  ${\vec{S}=\vec{S}_\alpha+\vec{S}_\kappa}$. However, this approach is far less accurate than solving the full matrix equation. In extreme tests with electron densities of $n_{\mathrm{CR}} > 10^{5}\ \mathrm{cm^{-2}}$ and magnetic fields of $B>5\ \mathrm{mG}$ the alternative solver scheme starts to kick in and we can get errors up to $5\ \% - 10\ \%$ per cell. However, we consider this error range still to be acceptable as long as the number of grid cells with extreme conditions is sufficiently small enough compared to the total number of grid cells. Furthermore, electron densities up to $10^{5}\ \mathrm{cm^{-2}}$ and a field strength in the order of $\mathrm{mG}$ are rather untypical ISM conditions.

As a last fail-save we skip certain cells completely and jump to the next one when the amount of required RKF45 solver steps exceeds $10^{7}$. Splitting of RT matrices and limiting the maximal amount of solver steps  allows the {\sc POLARIS} code to terminate in any case and within a reasonable time frame. We note, that none of these implemented fail-saves kick in for the Milky Way model of \citep[][]{ReisslA} utilised in this paper.

\section{An idealised toroidal field model}
\label{app:Toro}
\begin{figure}
\begin{center}
	\begin{minipage}[c]{0.65\linewidth}
	\begin{center}
	      \includegraphics[width=1.0\textwidth]{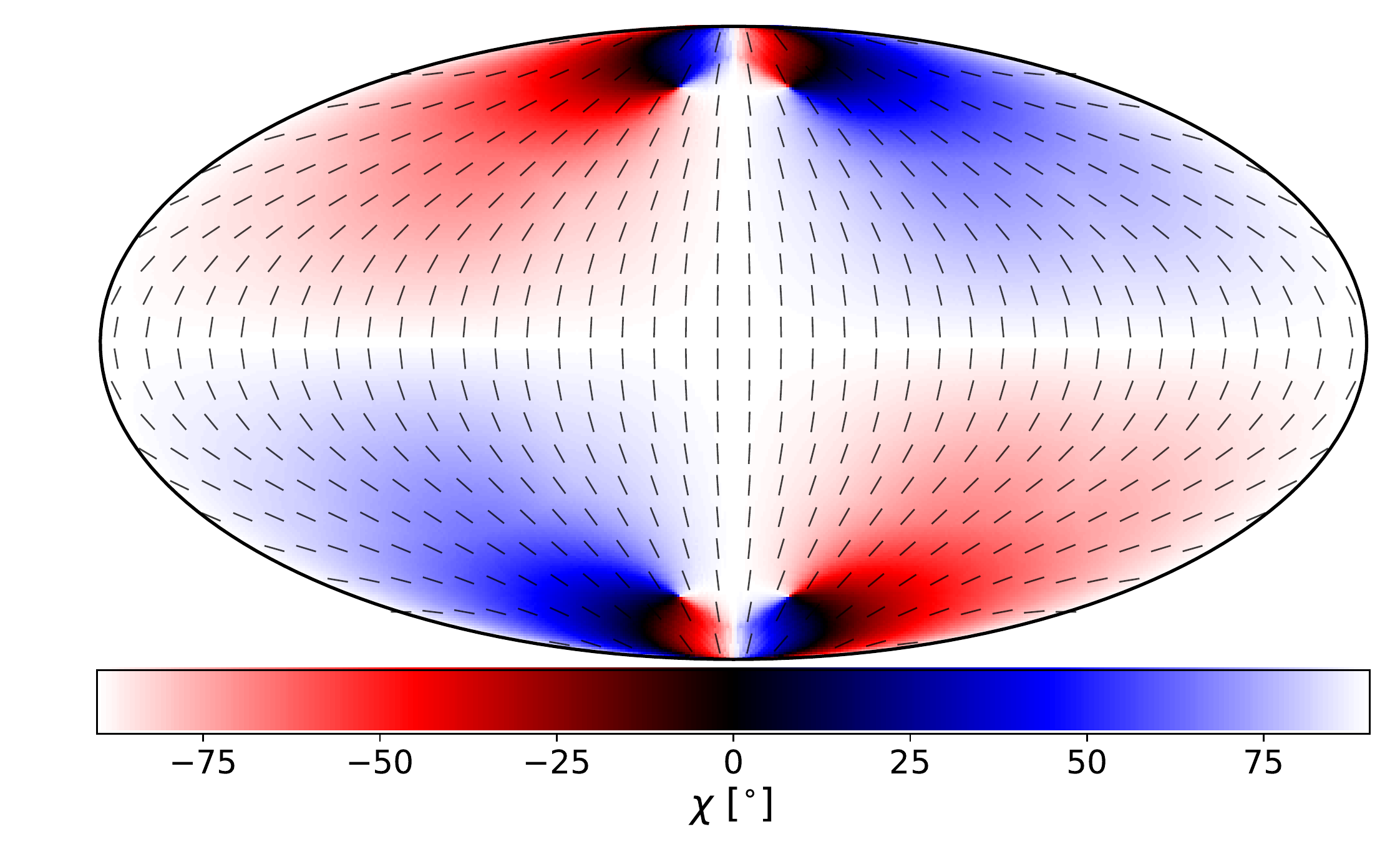}
	\end{center} 
	
	\caption{All-sky healpix projection of a purely toroidal magnetic field.}
	\label{fig:Toro}
	\end{minipage}
\end{center}
\end{figure}

Both, observations as well as modeled synchrotron all-sky polarisation maps discussed in this paper appear to be strongly controlled by a toroidal magnetic field component. Hence, we provide a polarisation map of a purely toroidal field in Figure \ref{fig:Toro} projected on a healpix sphere for comparison. The map is created with {\sc POLARIS} assuming constant densities and a perfectly polarised emission perpendicular to the field direction without any absorption. We note, that a purely toroidal field would posses a quadrupole-like symmetry with an orientation of the linear polarisation vectors close to $\pm 90^\circ$ along the galactic longitude as well as latitude.

\section*{Acknowledgements}
The authors thank the anonymous referee for valuable comments which improved the quality of the paper. Special thanks goes to Torsten En{\ss}lin, Alexander Pandya, and Jason Dexter for numerous enlightening discussions concerning synchrotron RT. We thank Juan Diego Soler for helpful conversations. We thank also the Auriga collaboration for generously sharing their data and the technical support. S.R., R.S.K., and E.W.P. acknowledge  support  from  the  Deutsche  Forschungsgemeinschaft in the Collaborative Research centre (SFB 881) ``The Milky Way System'' (subprojects B1, B2, and B8) and in the Priority Program SPP 1573 ``Physics of the Interstellar  Medium''  (grant  numbers  KL  1358/18.1,  KL  1358/19.2). The authors also acknowledge access to computing infrastructure support by the state of Baden-W\"urttemberg through bwHPC and the German Research Foundation (DFG) through grant INST 35/1134-1 FUGG.

\bibliographystyle{aasjournal}
\bibliography{bibtex}

\end{document}